\documentclass[fleqn,usenatbib]{mnras}
\usepackage[T1]{fontenc}
\usepackage{graphicx}	
\usepackage{amsmath}
\usepackage{amssymb}
\usepackage{ae,aecompl}
\usepackage{newtxtext,newtxmath}
\usepackage{soul}
\usepackage{comment}
\newcommand{\GG}[1]{}

\title[Origin of dwarf LSBGs]{The origin of low-surface-brightness galaxies in the dwarf regime}

\author[R. A. Jackson et al.]
{R. A. Jackson$^{1}$\thanks{E-mail: r.jackson9@herts.ac.uk},
G. Martin$^{2,3}$,
S. Kaviraj$^{1}$,
M. Rams{\o}y$^{4}$,
J. E. G. Devriendt$^{4}$, \newauthor 
T. Sedgwick$^{5}$,
C. Laigle$^{6}$, 
H. Choi$^{7}$,
R. S. Beckmann$^{6}$,
M. Volonteri$^{6}$,
Y. Dubois$^{6}$, \newauthor
C. Pichon$^{6,8}$, 
S. K. Yi$^{7}$, 
A. Slyz$^{4}$,
K. Kraljic$^{9}$,
T. Kimm$^{7}$,
S. Peirani$^{10}$ and
I. Baldry$^{5}$
\\
$^{1}$Centre for Astrophysics Research, School of Physics, Astronomy and Mathematics, University of Hertfordshire, Hatfield, AL10 9AB, UK\\
$^{2}$Steward Observatory, University of Arizona, 933 N. Cherry Ave, Tucson, AZ 85719, USA\\
$^{3}$Korea Astronomy and Space Science Institute, 776 Daedeokdae-ro, Yuseong-gu, Daejeon 34055, Korea\\
$^{4}$Dept of Physics, University of Oxford, Keble Road, Oxford OX1 3RH UK\\
$^{5}$Astrophysics Research Institute, Liverpool John Moores University, IC2, Liverpool Science
Park, 146 Brownlow Hill, L3 5RF\\
$^{6}$Institut d'Astrophysique de Paris, Sorbonne Universit\'es, UMPC Univ Paris 06 et CNRS, UMP 7095, 98 bis bd Arago, 75014 Paris, France\\
$^{7}$Department of Astronomy and Yonsei University Observatory, Yonsei University, Seoul 03722, Republic of Korea\\
$^{8}$School of Physics, Korea Institute for Advanced Study (KIAS), 85 Hoegiro, Dongdaemun-gu, Seoul, 02455, Republic of Korea\\
$^{9}$Institute for Astronomy, Royal Observatory, Edinburgh EH9 3HJ, UK\\
$^{10}$Observatoire de la C$\hat{\rm{o}}$te d'Azur, CNRS, Laboratoire Lagrange, Bd de l'Observatoire, Universit\'e C$\hat{\rm{o}}$te d'Azur, CS 34229, 06304 Nice Cedex 4, France
}

\pubyear{2020}

\begin{document}
\label{firstpage}
\pagerange{\pageref{firstpage}--\pageref{lastpage}}
\maketitle

\begin{abstract}
Low-surface-brightness galaxies (LSBGs) -- defined as systems that are fainter than the surface-brightness limits of past wide-area surveys -- form the overwhelming majority of galaxies in the dwarf regime (M$_{\star}$ $<$ 10$^9$ M$_{\odot}$). Using \texttt{NewHorizon}, a high-resolution cosmological simulation, we study the origin of LSBGs and explain why LSBGs at similar stellar mass show the large observed spread in surface brightness. \texttt{NewHorizon} galaxies populate a well-defined locus in the surface brightness -- stellar mass plane, with a spread of $\sim$3 mag arcsec$^{-2}$, in agreement with deep SDSS Stripe data. Galaxies with fainter surface brightnesses today are born in regions of higher dark-matter density. This results in faster gas accretion and more intense star formation at early epochs. The stronger resultant supernova feedback flattens gas profiles at a faster rate which, in turn, creates shallower stellar profiles (i.e. more diffuse systems) more rapidly. As star formation declines towards late epochs ($z<1$), the larger tidal perturbations and ram pressure experienced by these systems (due to their denser local environments) accelerate the divergence in surface brightness, by increasing their effective radii and reducing star formation respectively. A small minority of dwarfs depart from the main locus towards high surface brightnesses, making them detectable in past wide surveys (e.g. standard-depth SDSS images). These systems have anomalously high star-formation rates, triggered by recent, fly-by or merger-driven starbursts. We note that objects considered extreme/anomalous at the depth of current datasets, e.g. `ultra-diffuse galaxies', actually dominate the predicted dwarf population and will be routinely visible in future surveys like LSST.
\end{abstract}


\section{Introduction}
Our statistical understanding of galaxy evolution is fundamentally driven by objects that are brighter than the surface brightness limits of wide-area surveys. Rapid progress has been made over the last few decades in advancing our comprehension of how galaxies form and evolve over time. The observed multi-wavelength properties of galaxies have been mapped in detail, via large surveys from both ground and space-based instruments \citep[e.g.][]{Beckwith_2006,Bianchi2017,Blanton2017,Nayyeri2017,Aihara2019}. Confrontation of these surveys with simulations in cosmological volumes \citep[e.g.][]{Dubois2014,Vogelsberger2014,Schaye2015,Kaviraj2017} has enabled us to interpret this data and understand the physics of galaxy evolution. 

The current consensus from the recent literature is that galaxies form hierarchically, within a $\Lambda$CDM framework \citep[e.g.][]{Cole2000,Bullock2001,Hatton2003,Bower2006,Pipino2009}, with models based on this paradigm broadly reproducing the observed statistical properties of galaxies in contemporary surveys \citep[e.g.][]{Blanton2017,Aihara2019}. However, notwithstanding the successes of this standard model, our comprehension of galaxy evolution is largely restricted to relatively bright galaxies, that have effective surface brightnesses\footnote{The effective surface brightness is defined here as the mean surface brightness within the effective radius.} greater than the surface brightness limits of past wide surveys. For example, in surveys like the SDSS \citep{Abazajian2009}, which has provided much of the discovery space in the nearby Universe, galaxy completeness decreases rapidly at $\langle \mu\rangle_{e,r}$ > 23 mag arcsec$^{-2}$ \citep[e.g.][]{Cross2002,Blanton2005,Driver2005}, where $\langle \mu\rangle_{e,r}$ is the effective surface brightness in the $r$-band, dropping to $\sim$10 per cent for $\langle \mu\rangle_{e,r}$ $\sim$ 24 mag arcsec$^{-2}$ \citep[e.g][]{Kniazev2004}.

Thus, while our understanding of the evolution of \textit{bright} galaxies has progressed significantly, it is worth considering the significance of `low-surface brightness' galaxies (LSBGs), defined here as those that fall below the nominal surface brightness limits of past wide-area surveys and which are, therefore, undetectable in these datasets. Both theory \citep{Martin2019,Kulier2019} and observational work using small, deep surveys \citep[e.g.][]{McGaugh1995,Bothun1997,Dalcanton1997} indicate that most galaxies are, in fact, fainter than the surface brightness limits of past wide-area surveys. These LSBGs are a heterogeneous population, ranging from massive, diffuse disks to all dwarf galaxies at cosmological distances. Current cosmological simulations indicate that these LSBGs dominate the galaxy number density, comprising more than 85 per cent of objects down to M$_{\star}$ $\sim$ 10$^7$ M$_{\odot}$ \citep{Martin2019} and form a large, natural, empirically-unexplored extension of the population of bright galaxies on which our understanding of galaxy evolution is currently predicated.

The absence of these objects from past datasets has two important consequences. First, our empirical picture of galaxy formation is heavily biased. Second, since our models are statistically calibrated only to the subset of bright galaxies, our understanding of the physics of galaxy evolution remains potentially highly incomplete. It is, therefore, not surprising that many well-known tensions between theory and observation are in the low-surface brightness regime, e.g. the apparent excess of dwarfs in simulations i.e. the substructure problem \citep{Moore1999,Bullock2017}, the core-cusp problem \citep{Blok2010} and the `too-big-to-fail' problem \citep{Boylan-Kolchin2011}. A complete understanding of galaxy evolution therefore demands a detailed comprehension of how LSBGs evolve over cosmic time. 

LSBGs have seen an explosion of interest in the recent observational literature, driven by individual or small deep pointings \citep[e.g.][]{Kaviraj2014a,Kaviraj2014b,Mihos2015,Delgado2016,Merritt2016,Roman2017,Leisman2017,Greco2018,Kaviraj2019,Martin2020} and/or careful reprocessing of relatively shallow surveys to push fainter than their nominal detection limits \citep[e.g.][]{Trujillo2017,Sedgwick2019a}. While these efforts have started to give us a glimpse of the LSBG regime, the galaxy samples that underpin these studies are not representative of the general population of LSBGs. In particular, the statistical properties of LSBGs in groups  \citep[e.g][]{Castelli2016,Merritt2016,Roman2017b,Roman2017,Jiang2019} and the field \citep[e.g][]{MartinezDelgado2016,Papastergis2017,Leisman2017} remain particularly poorly understood, largely due to the lack of surveys that are both deep and wide. However, the successful detection of LSBGs in sparser environments indicates that they are not a cluster phenomenon and are, in fact, a ubiquitous population that inhabits all regions of the observable Universe. 

A burgeoning theoretical literature has explored the mechanisms that form LSBGs. For example, high halo spin \citep[e.g.][]{Amorisco2016}, bursty supernova feedback, which leads to the formation of cored dark matter haloes \citep[e.g.][]{DiCintio2017,Chan2018,Martin2019}, mergers \citep{Wright2020} or formation from high angular momentum gas \citep{Liao2019,Tremmel2019,DiCintio2019} have all been suggested as channels for creating LSBGs and `ultra-diffuse galaxies' (UDGs), which represent the extreme end of the LSBG population at the depth of current datasets. However, it has also been shown that environmental processes like tidal perturbations, the alignment of infalling baryons at early times and galaxy collisions are likely required to fully reproduce the variety and demographics of the LSBG/UDG populations seen in the observations \citep[e.g.][]{Baushev2018,Martin2019,Liao2019,Carleton2019,Tremmel2019,Cardona-Barrero2020,Jackson2020}. It is worth noting that LSBGs (including the more extreme UDG population) are predicted to form in all environments including the field \citep[e.g.][]{DiCintio2017,Chan2018,Jiang2019,Liao2019,Wright2020}, consistent with the findings of recent observational work.

A statistical comparison between observation and theory in the LSBG regime requires a hydrodynamical simulation in a cosmological volume. The hydrodynamics are essential for 2D predictions for baryons (which determines the surface brightness of the mock galaxies), while a cosmological volume is required for making statistical predictions for the properties of the LSBGs as a whole, across the full spectrum of cosmological environments (field, groups etc). In recent work, \citet{Martin2019} have performed a comprehensive study of the formation of relatively massive LSBGs, using the Horizon-AGN cosmological simulation \citep{Kaviraj2017}. They showed that, in the stellar mass range M$_{\star}$ $>$ 10$^9$ M$_{\odot}$, the formation of LSBGs and their eventual divergence from their high surface brightness counterparts, is triggered by a period of more intense star formation activity in the early ($z>2$) Universe. This leads to more intense supernova feedback, which moves gas from the central regions towards the outskirts, flattening their gas profiles, but typically does not remove gas completely from the system. These shallower gas profiles then lead to shallower stellar profiles which are more susceptible to tidal processes and ram pressure stripping. Over time, these processes `heat' the stellar and gas content of the LSBGs, increasing their effective radii further, and quench the galaxies, both of which lead to their low surface brightnesses at the present day. %\citet{Martin2019} have shown that, at least in the mass range that they study (M$_{\star}$ $>$ 10$^{9}$ M$_{\odot}$), LSBGs are not formed as a result of a high halo spin or formation from high angular momentum gas \citep[e.g.][]{Yozin2015,Amorisco2016,Rong2017,Amorisco2018}.

While \citet{Martin2019} has offered key insights into the formation of relatively massive LSBGs, the range of stellar masses that can be probed by Horizon-AGN (and other simulations with similar box sizes such as EAGLE \citep{Schaye2015} and Illustris \citep{2014MNRAS.444.1518V}) is limited by both its stellar mass resolution (M$_{\star}$ $\sim$ 10$^{8.5}$ M$_{\odot}$) and spatial resolution ($\sim$1 kpc). The formation of lower-mass i.e. dwarf LSBGs, which is the regime in which most observational LSBG studies are focused, requires a cosmological simulation with much better mass and spatial resolution ideally in the tens of parsecs. Recall that the scale height of the Milky Way is $\sim$300 pc \citep[e.g.][]{Kent1991,Corredoira2002,McMillan2011}, so much higher spatial resolution is needed to properly resolve dwarfs.  

In this study, we use the \texttt{NewHorizon} cosmological hydro-dynamical simulation, which has stellar mass and maximum spatial resolutions of 10$^4$ M$_{\odot}$ and 40 pc respectively, to study the origin of low-mass LSBGs, drilling down deep into the dwarf regime. \texttt{NewHorizon} offers better mass and spatial resolution than any other simulation with a comparable volume, making it ideally suited for this exercise. Our aims are two fold. First, we study the surface brightness vs. stellar mass plane, for galaxies down to stellar masses of M$_{\star}$ $\sim$ 10$^{6.5}$ M$_{\odot}$, and compare the position of the main locus of galaxies to existing observational data. 
%We show that the concept of `downsizing', i.e. the hypothesis that lower-mass galaxies keep forming stars towards lower epochs, is largely an artefact of the most quenched low-mass galaxies being too faint to be detectable in past/current surveys.  
Second, we study how different processes (e.g. feedback from supernovae and active galactic nuclei (AGN), ram pressure, tidal perturbations and galaxy mergers) drive the origin of dwarf LSBGs and produce the large observed spread in galaxy surface brightness at fixed stellar mass. 

%We show that processes such as tidal processes, mergers and feedback from AGN are not the principal drivers of the position of galaxies in the surface brightness vs stellar mass plane. However, at fixed stellar mass, galaxies that have the lower surface brightnesses exhibit \textit{earlier} star formation driven by being in higher density environments where they can access relatively more gas. Similar to the findings of \citet{Martin2019}, this earlier star formation makes gas profiles shallower which then produce lower-surface brightness systems. The epoch of divergence between low and high surface brightness galaxies moaves towards lower redshift at progressively higher stellar mass. 
This paper is structured as follows. In Section \ref{sec:NH}, we briefly describe the \texttt{NewHorizon} simulation. In Section \ref{sec:galaxy_prop}, we study the properties of galaxies in the surface brightness vs. stellar mass plane in the nearby Universe. In Section \ref{sec:galaxy_evol}, we study how different processes contribute to the position of galaxies in the surface brightness vs. stellar mass plane at low redshift. In Section \ref{sec:off_locus_section}, we explore why a minority of galaxies depart strongly from the main locus of objects that hosts the majority of galaxies in this plane. We summarise our findings in Section \ref{sec:summary}.

%.........................................................

\section{Simulation}
\label{sec:NH}

We use the \texttt{NewHorizon} cosmological, hydro-dynamical simulation \citep{Dubois2020} \footnote{\href{[http://new.horizon-simulation.org}{new.horizon-simulation.org}}, which is a high-resolution zoom of a region within the Horizon-AGN simulation \citep[][H-AGN hereafter]{Dubois2014,Kaviraj2017}. The simulation has been run down to $z=0.25$. NewHorizon employs the adaptive mesh refinement code RAMSES \citep{Teyssier2002}. Initial conditions are taken from H-AGN, which utilises a grid that spans a 142 comoving Mpc volume, using 1024$^3$ uniformly-distributed cubic cells with a constant mass resolution, using \texttt{MPGrafic}. For \texttt{NewHorizon}, this grid is resampled at higher resolution (using 4096$^3$ uniformly-distributed cubic cells), with the same cosmology ($\Omega_m$=0.272, $\Omega_b$=0.0455, $\Omega_{\Lambda}$=0.728, H$_0$=70.4 km s$^{-1}$ Mpc$^{-1}$ and $n_s$=0.967 \citep{Komatsu2011}).

The high-resolution zoom has a volume of $\sim$(16 Mpc)$^3$, taken from an average density region of H-AGN. It has a dark-matter (DM) resolution of 10$^6$ M$_{\odot}$ (compared to 8$\times$10$^7$ M$_{\odot}$ for H-AGN), stellar mass resolution of 10$^4$ M$_{\odot}$ (compared to 2$\times$10$^6$ M$_{\odot}$ in H-AGN) and a maximum spatial resolution of 34 pc (compared to 1 kpc in H-AGN). This makes \texttt{NewHorizon} the simulation with the highest spatial and stellar mass resolution in a cosmological volume and an ideal tool with which to study the dwarf galaxy population. Note that, given that the zoom region used to create \texttt{NewHorizon} has an average density, this simulation does not contain high-density environments like clusters. 

% Refinement from level 13 down to level 21 is allowed only where the passive scalar reaches a value of 0.01, following two criteria: a quasi-Lagrangian criterion with a mass threshold corresponding to eight times the initial mass resolution and a super-Lagrangian criterion where cells are refined if the local gas Jeans length is smaller than the cell length, and if the gas density is half that of the threshold for star formation. The maximum level of refinement (21) is only reached at $z<1.5$. The refinement level is 20 at $1.5<z<4$, 19 at $4<z<9$ and 18 above at $z>9$, in order to maintain a physical resolution at an almost constant value of 35 pc. \textbf{Mention that \texttt{NewHorizon} is at $z \sim0.7$. Do this in the introduction as well.}\\§

% \textbf{Need to look at some of the technical parts of this section again.\\ }

%.........................................................

\subsection{Star formation and stellar feedback} 

Gas cools via the initial mixture of Hydrogen and Helium, which is progressively enriched by metals produced by stellar evolution \citep{Sutherland1993,Rosen1995}. We assume photoionized equilibrium, with an ambient UV background after the Universe is re-ionized at $z=10$ \citep{Haardt1996}. Star formation occurs in gas with a hydrogen number density greater than $n_{H}>$10 H cm$^{-3}$ and a temperature lower than 2$\times$10$^4$ K, following a Schmidt-Kennicutt relation \citep{schmidt1959,Kennicutt1998}. The efficiency depends on the local turbulent Mach number and virial parameter $\alpha$=2E$_k$/|$E_g$|, where E$_k$ is the turbulent energy of the gas and E$_g$ is the gas gravitational binding energy \citep{Kimm2017}. A probability of forming a star particle of mass M$_{\star,res}$=10$^4$ M$_{\odot}$ is drawn at each time step according to the Schmidt-Kennicutt law.

Each star particle represents a coeval collection of stars with different masses. 31 percent of the stellar mass of this star particle (corresponding to stars more massive than 6 M$_{\odot}$) is assumed to explode as Type II
supernovae, 5 Myr after its birth. The fraction is calculated using a Chabrier initial mass function, with upper and lower mass limits of 150 M$_{\odot}$ and 0.1 M$_{\odot}$ \citep{Chabrier2005}. 

Supernova (SN) feedback is modelled in the form of both energy and momentum, ensuring that the final radial momentum is accurately captured during the snowplough phase of the expansion \citep{Kimm2014}. Each supernova has an initial energy of 10$^{51}$ erg and a progenitor mass of 10 M$_{\odot}$. In addition, pre-heating of the ambient gas by ultraviolet radiation from young OB stars is taken into account, by augmenting the final radial momentum from supernovae following \citet{Geen2015}.

%.........................................................

\subsection{Supermassive black holes and black-hole feedback}

Supermassive black holes (SMBHs) are considered to be sink particles, which accrete gas and impart feedback to their local surroundings, according to some fraction of the rest-mass energy of the accreted material. SMBHs form in regions with gas density larger than the threshold of star formation, with a seed mass of 10$^4$ M$_{\odot}$. New SMBHs are not allowed to form at a distance less than 50 kpc from other existing black holes. A dynamical gas drag force is applied to the SMBHs \citep{Ostriker1999} and two SMBHs are allowed to merge if the distance between them is smaller than 4 times the cell size, and if the kinetic energy of the binary is less than its binding energy. 

Black holes (BHs) accrete at the Bondi-Hoyle-Lyttleton accretion rate, with its value capped at Eddington \citep{hoyle1939,Bondi1944}. They release energy back into the gas, both via a jet `radio' mode and a thermal quasar mode, for accretion rates below and above 1 percent of the Eddington rate respectively \citep{Dubois2012}. SMBH spins are evolved self-consistently through gas accretion in the quasar mode and coalescence of black hole binaries \citep{Dubois2014b}. This modifies the radiative efficiencies of the accretion flow, following the models of thin Shakura \& Sunyaev accretion discs, and the corresponding Eddington accretion rate, mass-energy conversion, and bolometric luminosity of the quasar mode \citep{Shakura1973}. The quasar mode imparts a constant 15 percent of the bolometric luminosity as thermal energy back into the surrounding gas, while the radio mode has a spin-dependent variable efficiency and a spin up/down rate that follows results from simulations of magnetically choked accretion discs \citep{Mckinney2012}.

%.........................................................

\subsection{Selection of galaxies and construction of merger trees}

DM halos are identified using the AdaptaHOP algorithm \citep{Aubert2004,Tweed2009}, which efficiently removes subhalos from main structures and counts the fractional number of low-resolution DM particles within the DM virial radius. Galaxies are identified in a similar fashion using the HOP structure finder applied directly on star particles \citep{Eisenstein1998}. The difference with AdaptaHOP lies in the fact that HOP does not remove substructures from the main structure, since this would result in star-forming clumps being removed from galaxies. %However, satellites are included in main halos once they spatially connect (in the HOP sense) with their central.
We produce merger trees for each galaxy in the final snapshot ($z=0.25$), with an average timestep of $\sim$ 15 Myr, which allows us to track the main branch progenitors of each galaxy with high temporal resolution. 

Given that \texttt{NewHorizon} is a high resolution zoom of H-AGN, we also need to consider the DM purity of galaxies. It is possible for higher mass DM particles to enter the high resolution region of \texttt{NewHorizon} from the surrounding lower-resolution regions and, given the large mass difference, interact in unusual ways with the galaxies that they encounter. The vast majority of galaxies affected by low DM purity exist at the outer edge of the \texttt{NewHorizon} sphere. In the analysis that follows, we only consider galaxies with DM halos that are more than 99 per cent pure. We note that our results remain unchanged if we use the sample that is 100 per cent pure. We proceed with the 99 per cent pure sample because it puts our analysis on a firmer statistical footing.  

%In order to check that purity is not systematically affecting the galaxies, we created all the plots in the paper with a 100\% pure sample and find that contaminated galaxies inhabit all regions of the galaxy population and the main results are consistent. We therefore use the 99\% pure sample to increase the number of galaxies in the sample.

%.........................................................

\subsection{Local environment}
\label{sec:local_environment}

%To explore local environment we utilise two methods. First, following \citet{Martin2018a}, we define environment using the 3-D local number density of objects around each galaxy. Local density is calculated using an adaptive kernel density estimation method\footnote{The width of the kernel used for multivariate density estimation is responsive to the local density of the region, such that the error between the density estimate and the true density is minimised.} \citep{Breiman1977,Ferdosi2011,Martin2018a}. The density estimate takes into account all galaxies with stellar masses above 10$^{8}$ M$_{\odot}$. 

%Galaxies are then split into three percentile ranges in local density: `low density' corresponds to galaxies in the 0th -- 40th density percentiles, `intermediate density' correspond to the 40th -- 90th percentiles and `high density' corresponds to galaxies in the 90th -- 100th percentiles.

In some of our analysis below we explore details of the environment of galaxies in the cosmic web (e.g. their distances to nodes and filaments), using the persistence-based filament tracing algorithm \texttt{DisPerSE} \citep{Sousbie2011}, which uses the density field computed via a delaunay tessellation \citep{SchappetVandeWeygaert2000} of the DM halo distribution. We choose a persistence of 4 sigma. \texttt{DisPerSE} identifies ridges in the density field as special lines that connect topologically robust pairs of nodes. These ridges compose the filament network of the cosmic web, and the set of all segments defining these ridges are referred to as the `skeleton' \citep{Pogosyan2009}. The distance to the nearest filament and node is computed for each DM halo to form a filament catalog. We refer readers to \citet{Sousbie2011} and \citet{Sousbie2011b} for more details of the \texttt{DisPerSE} algorithm and to \citet{Dubois2014} and \citet{Laigle2018} for an implementation of this algorithm on H-AGN. 

%.........................................................

\subsection{Calculation of surface brightness}
\label{sec:surface_brightness_calc}

We obtain the intrinsic (i.e. unattenuated) surface brightness for each galaxy using the intensity-weighted central second-moment of the stellar particle distribution \citep[e.g.][]{Bernstein2002}. We calculate the surface brightness in multiple orientations ($xy$, $xz$ and $yz$) and use the mean value in our study. The procedure for calculating the surface brightness of individual galaxies is as follows.

We first obtain the intrinsic $r$-band magnitudes for each star particle that makes up the galaxy. To do this we obtain the full spectral energy distribution (SED) from a grid of \citet[][BC03 hereafter]{Bruzual2003} simple stellar population (SSP) models corresponding to the closest age and metallicity of each star particle. We redshift each BC03 template to the redshift of the galaxy and convolve the redshifted BC03 templates with the response curve for the SDSS $r$-band filter. We then weight by the particle mass to obtain the luminosity contribution of each star particle, and obtain the apparent $r$-band magnitude by converting the flux to a magnitude and adding the distance modulus and zero point.

We then obtain the second moment ellipse as follows. We first construct the covariance matrix of the intensity-weighted central second-moment for the star particles,

\begin{equation}
    \mathrm{cov}[I(x,y)]=\begin{bmatrix}
        Ix^{2}&Ixy\\
        Ixy&Iy^{2}\\
    \end{bmatrix},
\end{equation}

\noindent where $I$ is the flux of each star particle and $x$ and $y$ are the projected positions from the barycentre in arc seconds. We obtain major ($\alpha=\sqrt{\lambda_{1}/\Sigma I}$) and major ($\beta=\sqrt{\lambda_{2}/\Sigma I}$) axes of the ellipse from the covariance matrix, where $\lambda_{1}$ and $\lambda_{2}$ are its eigenvalues and $\Sigma I$ is the total flux, and find the scaling factor, $R$, which scales the ellipse so that it contains half the total flux of the object. Finally, we calculate the mean surface brightness within the effective radius, $\langle\mu\rangle_{e,r} = m - 2.5\mathrm{log_{10}}(2) + 2.5\mathrm{log_{10}}(A)$, where $A=R^{2} \alpha  \beta \pi$ and $m$ is the $r$-band apparent magnitude of the object.

\begin{figure*}
\centering
\includegraphics[width=\textwidth]{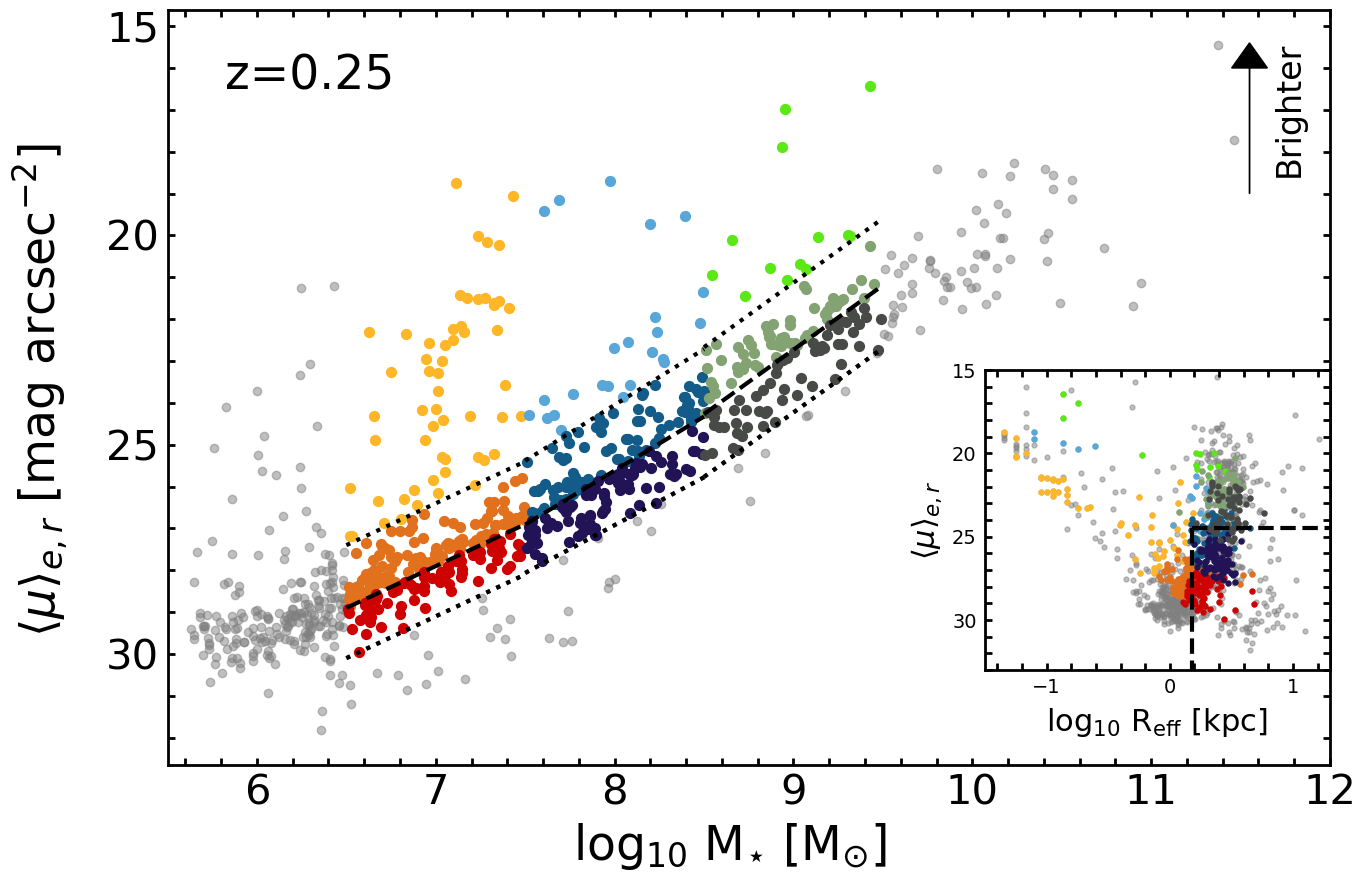}
\caption{Intrinsic effective surface brightness vs. stellar mass for galaxies in the \texttt{NewHorizon} simulation. The bulk of the galaxy population resides on a well-defined locus with a large spread of $\sim$3 dex, with a minority ($\sim$10-20 per cent, depending on the stellar mass range in question) departing strongly from the main locus towards higher surface brightness. The dotted lines indicate the limits of the main locus, defined by eye. The coloured points indicate different zones of the locus (described in Section \ref{sec:galaxy_prop}) which we consider for our analysis. The inset shows the effective radius vs. intrinsic effective surface brightness for the \texttt{NewHorizon} galaxies. The dashed lines in the inset indicate the typical boundaries that demarcate `ultra-diffuse galaxies' (UDGs) in the literature (R$_{\rm{eff}}$ > 1.5kpc and  $\langle \mu\rangle_{e,r}$ > 24.5). While UDGs are sometimes considered extreme or anomalous at the depth of current datasets, they actually dominate the predicted dwarf population, and will be routinely visible in future surveys like LSST.}
\label{fig:mvssurface brightness}
\end{figure*}

%.........................................................

\section{The surface brightness vs. stellar mass plane in the nearby Universe}
\label{sec:galaxy_prop}

Figure \ref{fig:mvssurface brightness} shows the intrinsic surface brightness vs. stellar mass plane for galaxies in \texttt{NewHorizon} at $z=0.25$. The majority of galaxies populate a locus in this plane, with a large spread of $\sim3$ mag arcsec$^{-2}$. A small minority of galaxies scatter off this locus towards high surface brightnesses. In order to understand the origin of galaxies that reside in different parts of the surface brightness vs. stellar mass plane at $z=0.25$, we split our galaxies into three mass bins. The `low', `intermediate' and `high' mass bins cover the mass ranges 10$^{6.5}$ M$_{\odot}$ $\geq$ M$_{\star}$ $>$ 10$^{7.5}$ M$_\odot$, 10$^{7.5}$ M$_\odot$ $\geq$ M$_{\star}$ $>$ 10$^{8.5}$ M$_{\odot}$ and 10$^{8.5}$ M$_{\odot}$ $\geq$ M$_{\star}$ $>$ 10$^{9.5}$ M$_{\odot}$ respectively. 

We define the limits of the main locus of objects by eye as the area where galaxies are closely clustered together. This is indicated by the dotted lines in Figure \ref{fig:mvssurface brightness}. Within each mass bin, we then split the galaxy population into three zones. We use a straight line that bisects the region between the two dotted lines in each mass bin to define the `lower' and `upper' zones, with galaxies brighter than the upper dotted line classified as `off-locus'. Note that surface brightness increases towards the upper end of Figure \ref{fig:mvssurface brightness}. Thus, the upper locus galaxies represent the population that is brighter in surface brightness at $z=0.25$, while the lower locus galaxies represent their fainter counterparts. These zones and mass bins are indicated using the different colours in Figure \ref{fig:mvssurface brightness}, where colour designates the mass bin and the shade represents the position of the galaxy (lower, upper or off) on the locus. Due to the small number of galaxies, and since this regime was comprehensively explored in \citet{Martin2019}, we do not study objects at the highest stellar masses (M$_{\star}$ $>$ 10$^{9.5}$ M$_\odot$). We also do not study galaxies which have M$_{\star}$ $<$ 10$^{6.5}$ M$_{\odot}$ because their progenitors are not massive enough to be well-resolved at early epochs.

The inset in Figure \ref{fig:mvssurface brightness} shows the intrinsic surface brightness vs. effective radius for the \texttt{NewHorizon} galaxy population. The dashed lines indicate the typical values that are used to identify `ultra-diffuse galaxies' \citep[UDGs; e.g.][]{Koda2015,Dokkum2015,Conselice2018,Laporte2019}, which form the faint end of the LSBG population that is detectable at the surface brightness limits of past/current datasets. These systems are sometimes considered to be potentially extreme or anomalous, due to their low surface brightnesses and extended nature. A variety of formation mechanisms have been proposed for their formation, such as the puffing up of `normal' dwarfs due to internal feedback processes \citep[e.g.][]{Amorisco2016,DiCintio2017} and the possibility that UDGs (particularly those in clusters) may be `failed' galaxies with anomalously low star formation efficiencies \citep[e.g.][]{Dokkum2015udg,Dokkum2015udgb,Koda2015}. 

\begin{figure*}
\centering
\includegraphics[width=\textwidth]{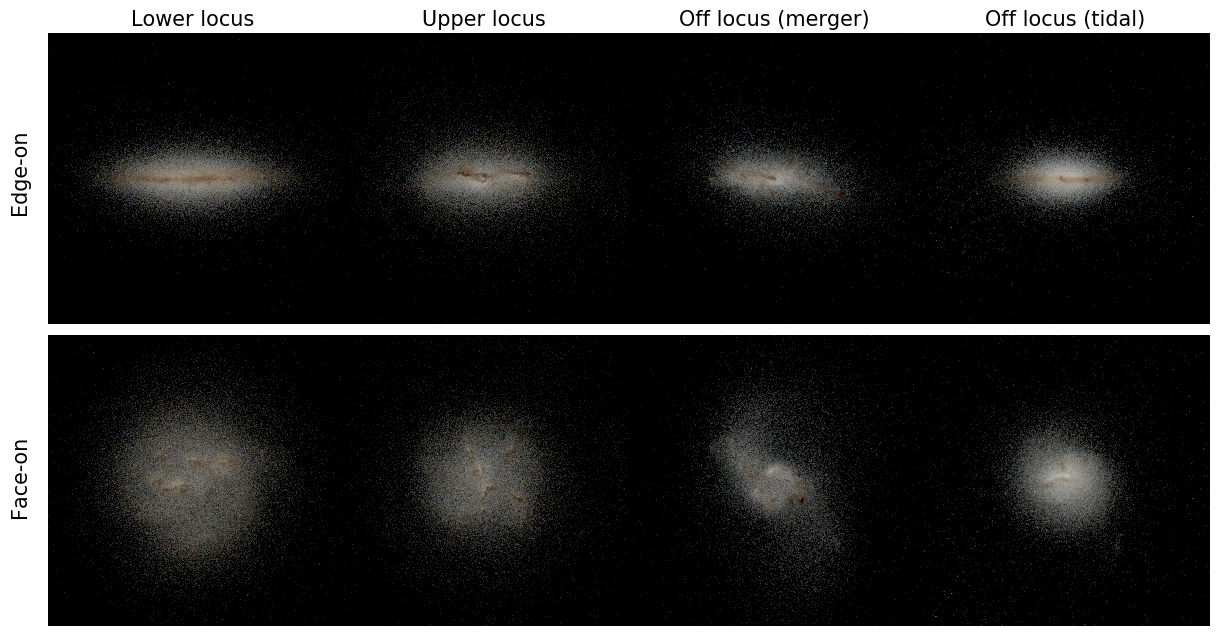}
\caption{Mock $gri$ images of dwarf galaxies in the \texttt{NewHorizon} simulation, created using the SKIRT9 code. Each column shows a typical galaxy in one of the zones on the main locus of objects -- a lower and upper locus galaxy and two off-locus systems produced via tidal perturbations and mergers respectively (see Section \ref{sec:off_locus_section} for an exploration of how off-locus galaxies form). The top and bottom rows show edge-on and face-on projections for each galaxy respectively. Galaxies in \texttt{NewHorizon} tend to become more compact as they move towards higher surface brightness (i.e. as we move from lower to off-locus systems).}
\label{fig:images}
\end{figure*}

However, the inset in Figure \ref{fig:mvssurface brightness} indicates that a large number of dwarf galaxies in \texttt{NewHorizon}, particularly in the low and intermediate mass bins, have surface brightnesses and effective radii that make them consistent with the definition of UDGs in the observational literature. UDGs are, therefore, a normal component of the dwarf galaxy population at low surface brightnesses. Note that since \texttt{NewHorizon} does not contain any clusters, a clear prediction is that UDGs should exist in large numbers in groups and the field and should be routinely detectable in new and future deep surveys such as the Hyper Suprime-Cam Subaru Strategic Program \citep[HSC-SSP][]{Aihara2019} and the Legacy Survey of Space and time (LSST) on the Vera C. Rubin Observatory \citep{Robertson2019}. In Figure \ref{fig:images} we present mock images of typical galaxies that occupy different regions of the locus, created using the \texttt{SKIRT9} code \citep{Baes2020}, which employs full radiative transfer based on the stars and gas within a galaxy. 

\begin{figure}
\centering
\includegraphics[width=\columnwidth]{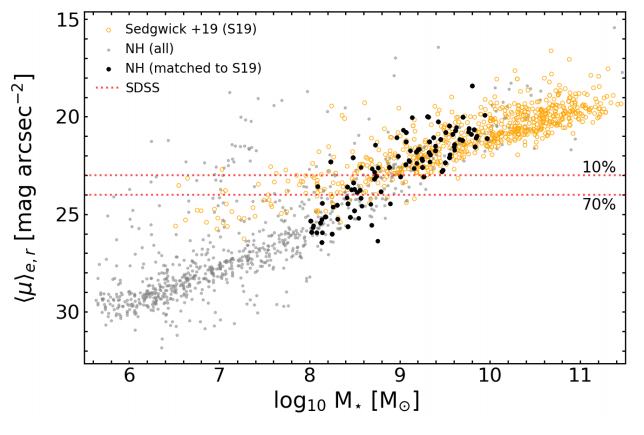}
\caption{Surface brightness (including dust attenuation) vs. stellar mass in the \texttt{NewHorizon} simulation. Grey points indicate the entire galaxy population of \texttt{NewHorizon}, open yellow points indicate galaxies from \citet{Sedgwick2019b} and black points are \texttt{NewHorizon} galaxies that are selected to match the 2D $M_{\star}-z$  distribution of \citet{Sedgwick2019b}. The predicted surface brightness vs. stellar mass plane in \texttt{NewHorizon} corresponds well to that in the observations, in the stellar mass range where the Sedgwick et al. galaxies are complete (recall that the simulation is not calibrated to produce galaxy surface brightnesses). The red dotted lines indicate the 70$\%$ and 10$\%$ completeness limits from the SDSS \citep[see e.g. Table 1 in][]{Blanton2005}. The detectability of galaxies in benchmark wide area surveys like the SDSS decreases rapidly in the dwarf galaxy regime. For example, at M$_{\star}$ < 10$^8$ M$_{\odot}$, the overwhelming majority of galaxies in the Universe lie below the surface brightness thresholds of surveys like the SDSS, with only those galaxies that depart strongly from the locus likely to be detectable in these datasets.}
\label{fig:locus_comparison}
\end{figure}

While we study the galaxy population in terms of intrinsic surface brightness for our analysis in Sections \ref{sec:galaxy_evol} and \ref{sec:off_locus_section}, we use the attenuated surface brightness for our comparison to observations below. In Figure \ref{fig:locus_comparison} we compare the predicted surface brightness vs. stellar mass plane in \texttt{NewHorizon} to that from recent work that uses the SDSS Stripe 82, which is $\sim$2 mags deeper than standard-depth SDSS imaging \citep{Sedgwick2019b,Sedgwick2019a}. Sedgwick et al. use a novel technique that allows them to identify LSBGs that do not appear in the pipeline-constructed galaxy catalogue in the IAC Stripe 82 Legacy Project \citep{Fliri2016} because they are too faint. They achieve this by identifying a sample of galaxies using core-collapse supernovae (CCSNe). Using custom settings in SExtractor \citep{bertin1996sextractor} they then extract the host galaxies of these CCSNe, many of which are not detected by the original SDSS Stripe 82 pipeline or by the IAC Stripe 82 legacy survey. The resultant sample is free of incompleteness in surface brightness in the stellar mass range M$_{\star}$ > 10$^8$ M$_{\odot}$, with a host being identified for all 707 CCSNe candidates at $z<0.2$. Of this sub-sample, 251 are spectroscopically confirmed CCSNe, with the remainder classified from the shape of their light-curves \citep[see][]{Sako2018}. This sample is well-suited to a comparison with \texttt{NewHorizon}, both due to its high completeness at low surface brightness and also because we can relatively easily model the selection function and apply it to our simulated dataset. 

Since the detectability of the \citet{Sedgwick2019a} galaxies depends on them hosting CCSNe, and therefore hosting star formation activity, we restrict our comparison to a subset of star-forming \texttt{NewHorizon} galaxies. The associated selection probability is correlated with the star formation rate (SFR), because the SFR determines the rate of CCSNe. We create a matching sample to the one from \citet{Sedgwick2019a} by drawing galaxies with a weight proportional to the normalised probability distribution in SFR. For the comparison in the $M_{\star}-\langle \mu \rangle_{\rm{e}}$ plane, we calculate the surface brightness of each \texttt{NewHorizon} galaxy as described in Section \ref{sec:surface_brightness_calc}. We additionally implement dust attenuation via a screen model using the SUNSET code \citep[see][]{Kaviraj2017}. Figure \ref{fig:locus_comparison} shows that the predicted surface brightness vs. stellar mass plane in \texttt{NewHorizon} corresponds well to that in the observations in the stellar mass range where the Sedgwick et al. galaxies are complete (recall that the simulation is not calibrated to produce galaxy surface brightnesses). The flattening seen in the observations is due to high levels of incompleteness at M$_{\star}$ < 10$^8$. The good overlap between the \texttt{NewHorizon} galaxies and the Sedgwick et al. data suggests that, while the prescriptions used to describe baryonic processes in \texttt{NewHorizon} (e.g. SN feedback) are largely tuned to higher mass galaxies, they offer a reasonable representation of these processes also in the low-mass regime.

Finally, it is worth noting that the surface-brightness limits of many past benchmark wide-area surveys lead to high levels of incompleteness in the low-mass regime. The dotted lines in Figure \ref{fig:locus_comparison} indicate the galaxy completeness limits for standard-depth SDSS imaging in the local Universe \citep{Blanton2005}. Note that no evolution correction is applied to modify these completeness limits for $z=0.25$ (the redshift of the \texttt{NewHorizon} galaxies). At an effective surface brightness of 24 mag arcsec$^{-2}$ the galaxy completeness in SDSS is only $\sim$10 per cent. Indeed, the \texttt{NewHorizon} locus indicates that at M$_{\star}$ < 10$^8$ M$_{\odot}$, the overwhelming majority of galaxies in the nearby Universe are undetectable in surveys like the standard-depth SDSS, with \textit{only} those objects that depart strongly from the locus towards very high surface brightnesses likely to be present in such datasets at all. As we show in Section \ref{sec:off_locus_section}, these off-locus systems host anomalous levels of star formation triggered by a recent interaction, making them highly unrepresentative of the general galaxy population in these mass regimes. Therefore, caution needs to be exercised about drawing conclusions about galaxy evolution in general using low-mass galaxies that are detected in shallow, wide-area surveys on which our understanding of the extra-galactic Universe is currently predicated. It is worth noting that future deep-wide surveys like the HSC-SSP and LSST (which have limiting surface brightnesses of $\sim$31.5 mag arcsec$^{-2}$ in their deepest layers), should offer 100 per cent completeness for galaxies at least down to M$_{\star}$ $\sim$ 10$^{7}$ M$_{\odot}$ in the nearby Universe, providing a revolutionary increase in discovery space over past wide surveys like the SDSS. 

%.........................................................

\section{Galaxy evolution as a function of surface brightness: the impact of different processes}
\label{sec:galaxy_evol}

The analysis that follows studies the impact of different processes on the progenitors of galaxies in the different zones described in Figure \ref{fig:mvssurface brightness}, in order to explore how this main locus of objects forms over cosmic time. Recall that we split the locus into `low', `intermediate' and `high' mass bins that cover the mass ranges 10$^{6.5}$ M$_{\odot}$ $\geq$ M$_{\star}$ $>$ 10$^{7.5}$ M$_\odot$, 10$^{7.5}$ M$_\odot$ $\geq$ M$_{\star}$ $>$ 10$^{8.5}$ M$_{\odot}$ and 10$^{8.5}$ M$_{\odot}$ $\geq$ M$_{\star}$ $>$ 10$^{9.5}$ M$_{\odot}$ respectively. Each mass bin then split into upper, lower and off-locus zones. The upper and lower zones are defined by the brighter and fainter halves of the locus respectively and off-locus objects are those that scatter to higher surface brightnesses beyond the upper locus. In all the plots that follow, the colour coding of the median lines and error regions corresponds to the colours of the zones in Figure \ref{fig:mvssurface brightness}. Given the small number of off-locus galaxies in any mass bin we do not show error regions for this population for clarity. In this section, we focus on galaxies that reside on the main locus. We study the formation of off-locus galaxies in Section \ref{sec:off_locus_section}.

\begin{figure}
\centering
\includegraphics[width=\columnwidth]{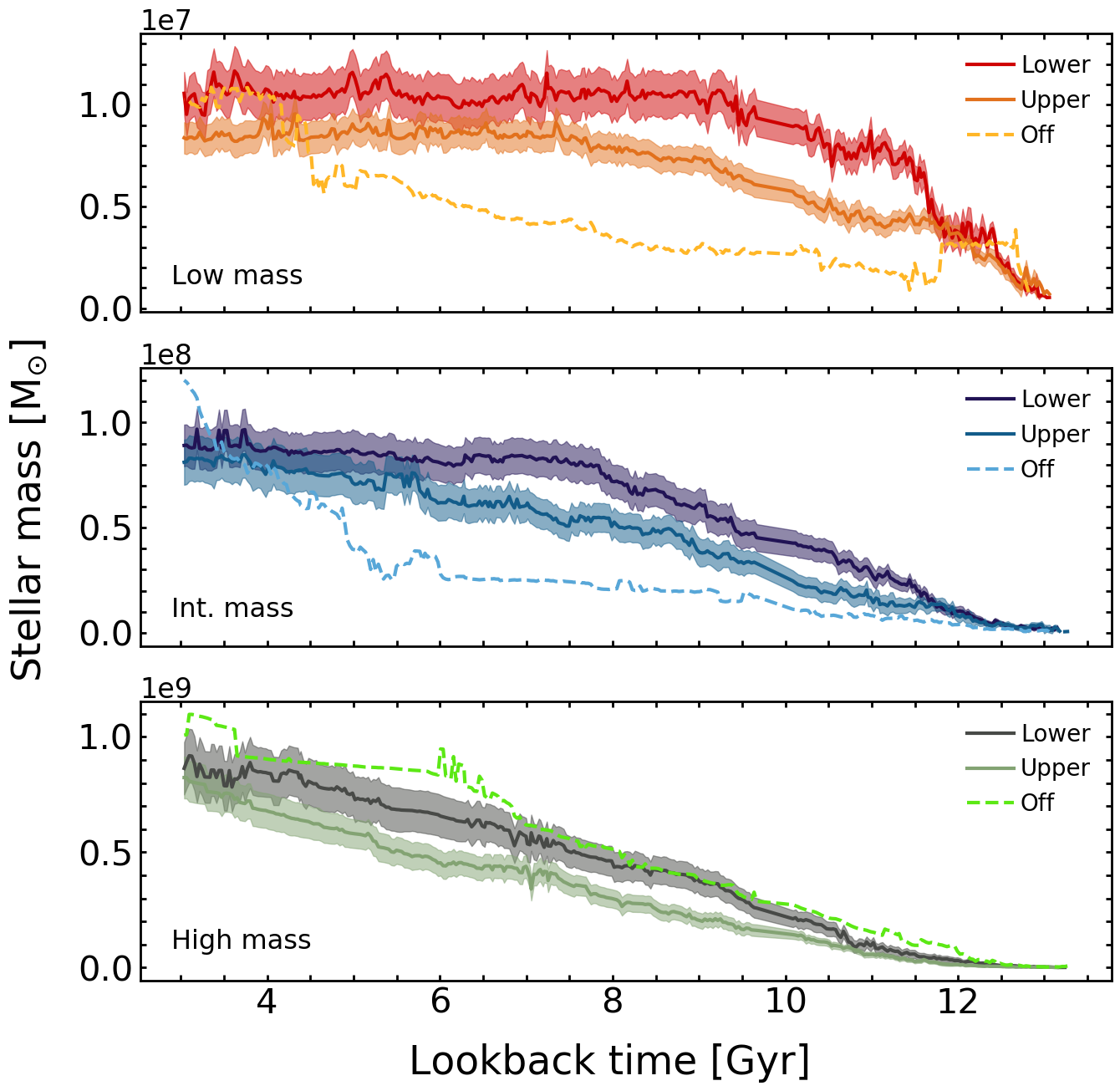}
\caption{Evolution of the median stellar mass with look-back time. The shaded regions show the associated uncertainties. The top, middle and bottom panels represent the low (10$^{6.5}$ M$_{\odot}$ $\geq$ M$_{\star}$ $>$ 10$^{7.5}$ M$_\odot$), intermediate (10$^{7.5}$ M$_\odot$ $\geq$ M$_{\star}$ $>$ 10$^{8.5}$ M$_{\odot}$) and high (10$^{8.5}$ M$_{\odot}$ $\geq$ M$_{\star}$ $>$ 10$^{9.5}$ M$_{\odot}$) mass bins respectively. The colour coding indicates the upper, lower and off locus galaxy populations. Recall that lower locus galaxies represent the population that is fainter in surface brightness at $z=0.25$, while the upper locus galaxies represent their brighter counterparts. Regardless of the mass bin in question, lower locus galaxies form their stars at earlier epochs than their upper locus counterparts. However, this stellar assembly bias takes place later at higher stellar masses.}
\label{fig:mass_evo}
\end{figure}

\begin{figure}
\centering
\includegraphics[width=\columnwidth]{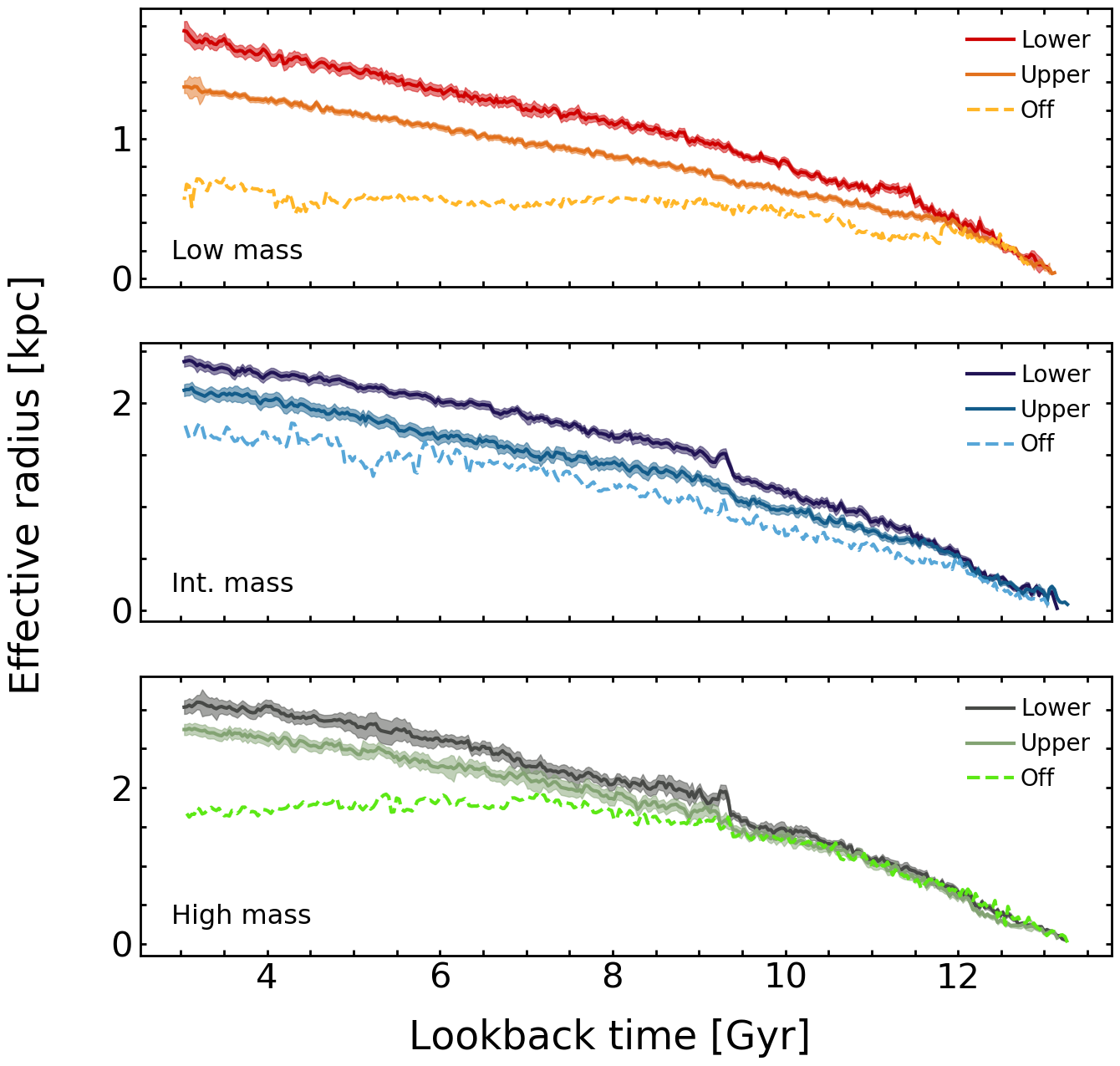}
\caption{Evolution of the median effective radius of galaxies with look-back time. The shaded regions show the associated uncertainties. The top, middle and bottom panels represent the low (10$^{6.5}$ M$_{\odot}$ $\geq$ M$_{\star}$ $>$ 10$^{7.5}$ M$_\odot$), intermediate (10$^{7.5}$ M$_\odot$ $\geq$ M$_{\star}$ $>$ 10$^{8.5}$ M$_{\odot}$) and high (10$^{8.5}$ M$_{\odot}$ $\geq$ M$_{\star}$ $>$ 10$^{9.5}$ M$_{\odot}$) mass bins respectively. The colour coding indicates the upper, lower and off locus galaxy populations. Recall that lower locus galaxies represent the population that is fainter in surface brightness at $z=0.25$, while the upper locus galaxies represent their brighter counterparts. In all mass bins, lower locus galaxies increase their effective radii faster than their upper locus counterparts.}
\label{fig:reff_evo}
\end{figure}

\begin{figure}
\centering
\includegraphics[width=\columnwidth]{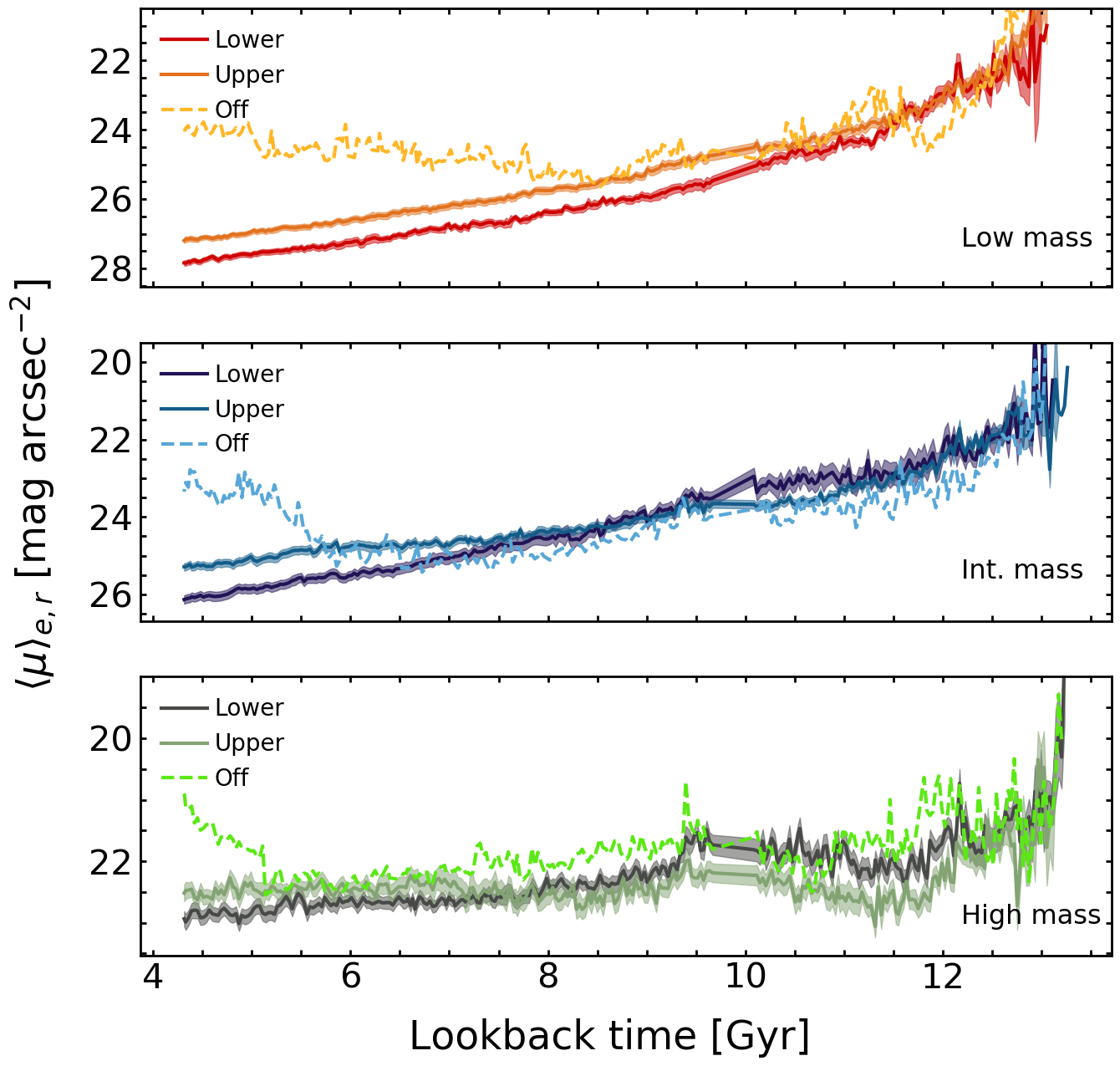}
\caption{Evolution of the median surface brightness with look-back time. The shaded regions show the associated uncertainties. The top, middle and bottom panels represent the low (10$^{6.5}$ M$_{\odot}$ $\geq$ M$_{\star}$ $>$ 10$^{7.5}$ M$_\odot$), intermediate (10$^{7.5}$ M$_\odot$ $\geq$ M$_{\star}$ $>$ 10$^{8.5}$ M$_{\odot}$) and high (10$^{8.5}$ M$_{\odot}$ $\geq$ M$_{\star}$ $>$ 10$^{9.5}$ M$_{\odot}$) mass bins respectively. The colour coding indicates the upper, lower and off locus galaxy populations. Recall that lower locus galaxies represent the population that is fainter in surface brightness at $z=0.25$, while the upper locus galaxies represent their brighter counterparts. In all mass bins, lower locus galaxies exhibit similar surface brightnesses to their upper locus counterparts at early epochs because their earlier star formation counteracts their larger effective radii. The surface brightnesses eventually diverge, with the divergence occurring later at higher stellar masses.}
\label{fig:surface brightnessevo}
\end{figure}

We begin by exploring, in Figures \ref{fig:mass_evo}, \ref{fig:reff_evo} and \ref{fig:surface brightnessevo}, the evolution of the median stellar mass, effective radius and surface brightness of galaxies in different zones within the locus across cosmic time. At a given stellar mass, galaxies in the lower locus show a stellar assembly bias relative to those in the upper locus, in the sense that their stellar mass assembly takes place at earlier epochs, as shown in Figure \ref{fig:mass_evo}. The median effective radius of galaxies (Figure \ref{fig:reff_evo}) in the lower locus increases faster than their upper locus counterparts at all epochs, including the epoch where the stellar assembly bias is most pronounced. It is worth noting that starbursts typically add new stars to the central regions of galaxies and thus act to reduce the effective radius, opposite to the trend seen in Figure \ref{fig:reff_evo}. This suggests that, in the early Universe, the processes that drive the stellar assembly bias are also responsible for the differential evolution in median effective radius between the upper and lower locus galaxies. 

Figure \ref{fig:surface brightnessevo} shows the evolution of surface brightness for galaxies in different zones of the locus. The eventual divergence in surface brightness between the upper and lower locus populations (in all mass bins) is delayed compared to that in the stellar mass and effective radius, because the increase in radius is initially compensated for by the more rapid buildup of stellar mass. It is worth noting that the point at which the divergence in surface brightness takes place shows a dependence on stellar mass, with the divergence taking place later at higher stellar masses. This suggests that the depth of the potential well provides a stabilising influence against the processes that drive the surface brightness divergence between upper and lower locus galaxies. 

Both these points imply that the width of the locus i.e the surface brightness separation between the upper and lower locus galaxies is likely to be driven by the assembly history of galaxies, with those forming their stellar mass earlier residing at fainter surface brightnesses at the present epoch. Therefore, in order to explain the position of galaxies in the locus, we need to understand the processes responsible for the divergence in stellar mass and effective radius of the upper and lower locus populations. These processes can be either internal or external. Internal mechanisms include feedback from either SN or AGN (or both). External processes include tidal perturbations (including mergers) and ram pressure.  In the sections below, we consider how each of these processes affect galaxies, in order to identify the principal processes that determine the position of galaxies in the locus at $z=0.25$.

%.........................................................

\subsection{Differential supernova feedback at early epochs - initiating the divergence in surface brightness}

Given the coincidence of stellar assembly bias and the differential increase in the median effective radius of the upper and lower locus populations, we begin by considering the role of SN feedback in the surface brightness evolution of galaxies. Previous theoretical work has shown that, particularly at low stellar masses, where gravitational potential wells are shallow, SN feedback is capable of driving gas outflows \citep[e.g.][]{Dubois2008,DiCintio2017,Chan2018} which make the density profiles of the gas, and the stars that form from it, shallower \citep[e.g.][]{Martin2019}. The shallower stellar profiles naturally lead to more diffuse systems which have lower surface brightness. %There is also evidence that, through dynamical heating of stars, this process can increase a galaxy's effective radius \citep{Chan2015,El-Badry2016}. 

In order to calculate SN feedback we define the total mechanical and thermal energy released by stellar processes between two timesteps, $t_{0}$ and $t_{1}$, by summing the energy released by each star particle in a galaxy within this interval:

\begin{equation}
E_{\mathrm{SN}} = \sum_{i} m_{\odot,i} (E(z_{1})_{i}-E(z_{0})_{i}),
\end{equation}

\noindent where $m_{\odot,i}$ is the mass of an individual star particle and $E(z)_{i}$ is the cumulative mechanical and thermal energy released by that star particle, as a result of Type II SN, between the time of its formation and a redshift $z$.

\begin{figure}
\centering
\includegraphics[width=\columnwidth]{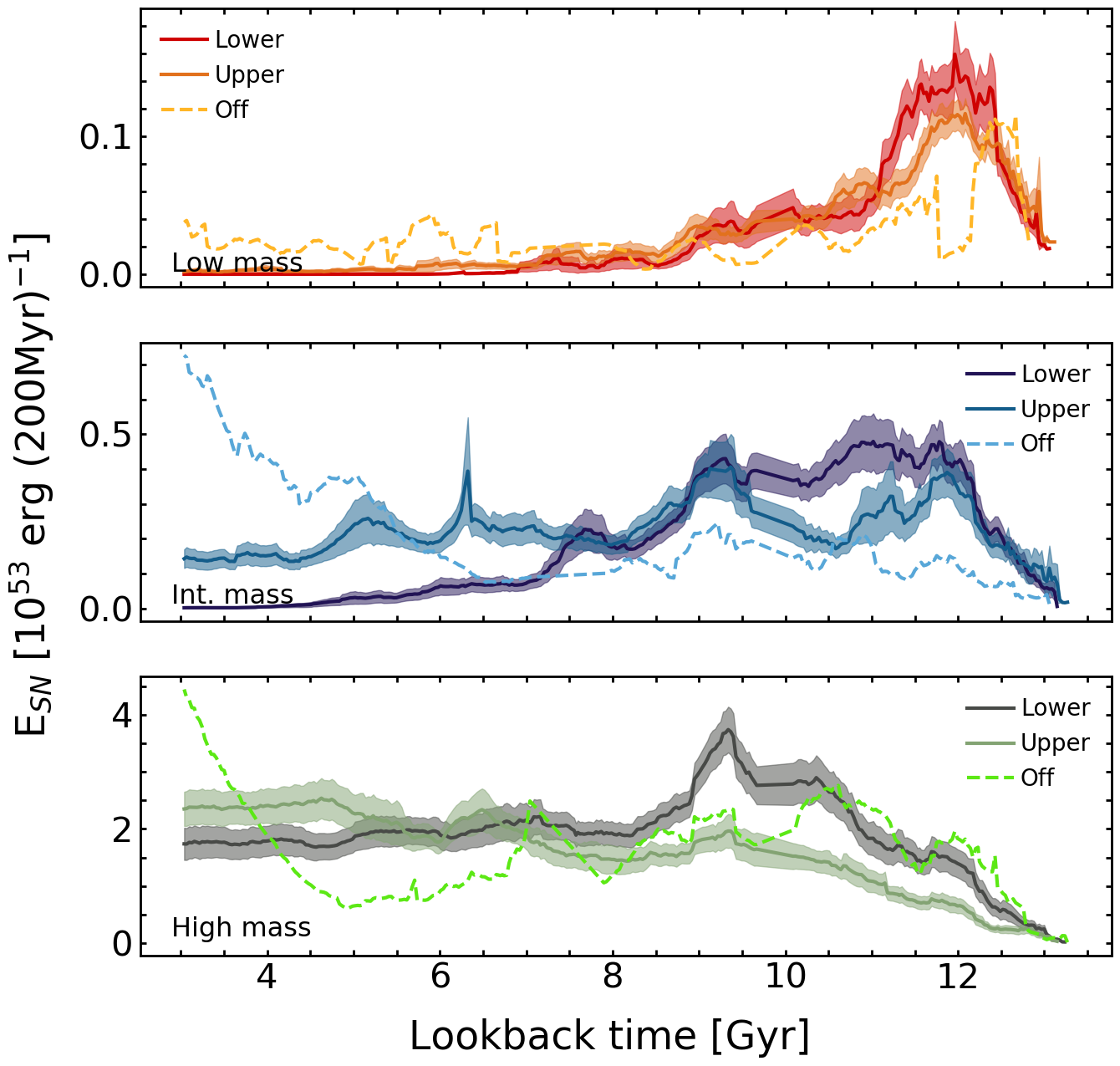}
\caption{Evolution of the median SN feedback as a function of look-back time. The shaded regions show the associated uncertainties. The top, middle and bottom panels represent the low (10$^{6.5}$ M$_{\odot}$ $\geq$ M$_{\star}$ $>$ 10$^{7.5}$ M$_\odot$), intermediate (10$^{7.5}$ M$_\odot$ $\geq$ M$_{\star}$ $>$ 10$^{8.5}$ M$_{\odot}$) and high (10$^{8.5}$ M$_{\odot}$ $\geq$ M$_{\star}$ $>$ 10$^{9.5}$ M$_{\odot}$) mass bins respectively. The colour coding indicates the upper, lower and off locus galaxy populations. Recall that lower locus galaxies represent the population that is fainter in surface brightness at $z=0.25$, while the upper locus galaxies represent their brighter counterparts. In all mass bins lower locus galaxies show higher levels of SN feedback (i.e. higher star formation rates) at earlier times, with the epoch of peak divergence occurring later at higher stellar masses. This drives the stellar assembly bias seen in all mass bins in Figure \ref{fig:mass_evo}.}
\label{fig:snevo}
\end{figure}

\begin{figure*}
\centering
\includegraphics[width=\textwidth]{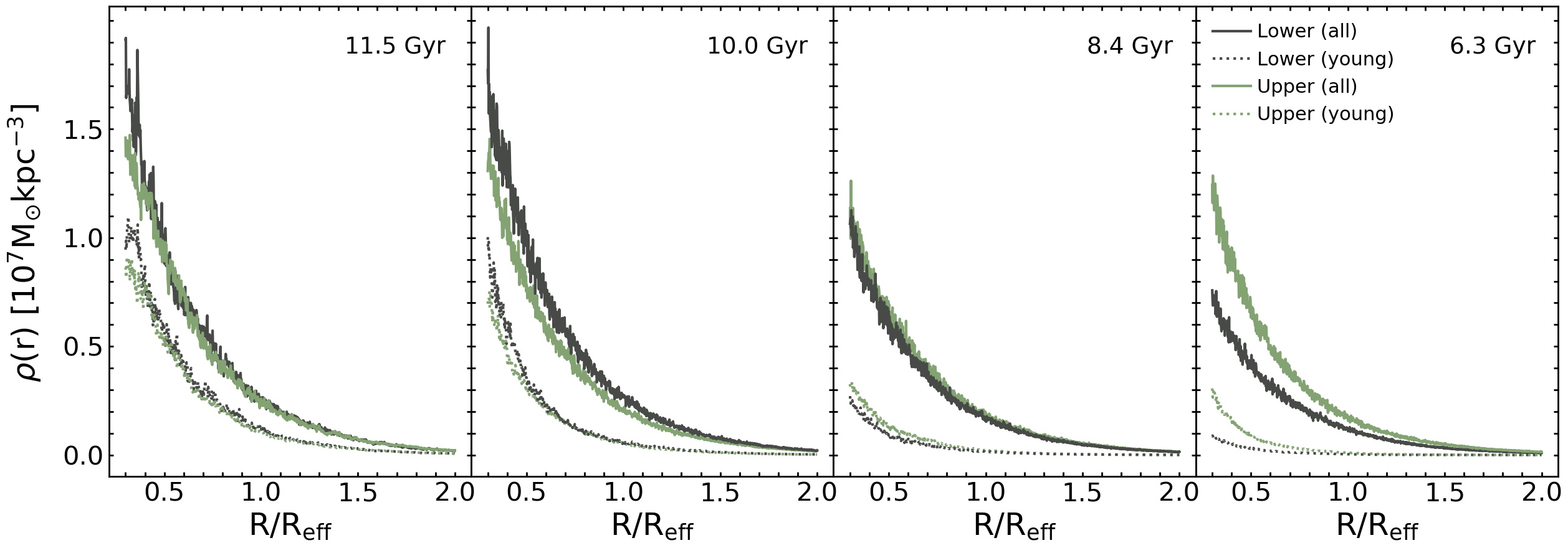}
\caption{Evolution of the stacked stellar density profiles of galaxies in the high mass bin as a function of look-back time, as indicated in the panels. The solid lines indicate all stars, while the dotted lines indicate young (ages < 100 Myr) stars only (which trace the star-forming gas). Dark and light green curves indicate lower and upper locus galaxies respectively. As SN feedback moves material away from the centres of galaxies, the density profiles of new stars become progressively flatter which, in turn, gradually flattens the overall stellar density profiles of galaxies. The higher SN feedback experienced by lower locus galaxies (see Figure \ref{fig:snevo}) at earlier times means that their slopes become flatter faster, as shown in the transition from the left to the right-hand panels above.}
\label{fig:prof}
\end{figure*}

In Figure \ref{fig:snevo}, we show the evolution of the SN feedback energy for galaxies in different zones of the locus. We find that galaxies which end up with fainter surface brightnesses (i.e. those in the lower locus) show higher levels of SN feedback at earlier times, regardless of the mass bin being considered. This correlates with the stellar assembly bias seen in Figure \ref{fig:mass_evo}. Note also that the stellar assembly bias occurs earlier at lower stellar masses, which is mirrored in the SN feedback being more pronounced at earlier epochs in lower mass bins.  

We proceed by exploring how this difference in SN feedback drives the evolution in the stellar mass, effective radius and surface brightness of galaxies shown in Figures \ref{fig:mass_evo}, \ref{fig:reff_evo} and \ref{fig:surface brightnessevo}. Recall that the stellar effective radii of upper and lower locus galaxies begin diverging in the early Universe (Figure \ref{fig:reff_evo}), around the epochs when there is a divergence in the SN feedback energy. This suggests that the SN feedback is likely to be impacting the gas reservoir in such a way as to drive the increase in the median effective radius.

For example, \citet{Martin2019} have shown that in the massive (M$_{\star}$ $>$ 10$^9$ M$_{\odot}$) galaxy regime, SN feedback acts to rapidly make the gas distribution shallower, by moving material from the central regions of the galaxy towards the outskirts \citep[see also][]{Governato2007}. If the gas is not completely removed, new stars forming from this gas make the stellar distribution progressively shallower over time. In this mass regime at least, earlier SN feedback is central to the formation of galaxies which exhibit lower surface brightnesses at the present day.

\begin{figure}
\centering
\includegraphics[width=\columnwidth]{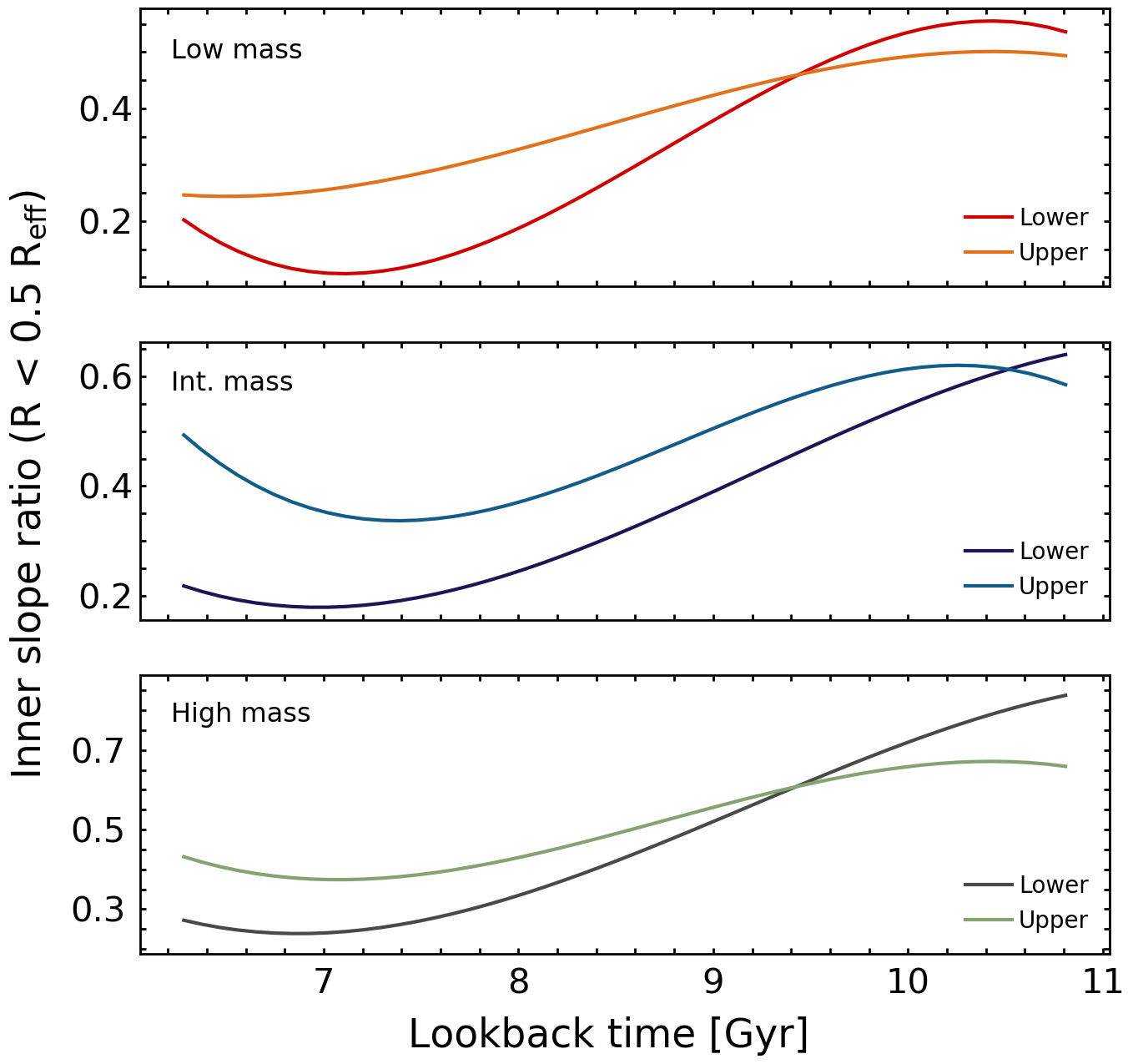}
\caption{Evolution of the ratio between the slope of the stellar density profiles of young and all stars, for R < 0.5 R$_{\rm{eff}}$. The slope is calculated by fitting a straight line to the inner 0.5 R$_{\rm{eff}}$ of the density profile. A value less than 1 indicates that the density profile of the young stars are flatter (i.e. shallower) than that of all stars. In lower locus galaxies, these ratios exhibit smaller values from early look-back times i.e. young stars have much flatter profiles compared to the overall stellar distribution, compared to upper locus galaxies. Since young stars will progressively flatten the entire stellar density distribution, this means that lower locus galaxies become more diffuse at a faster rate, as seen in the faster evolution of the effective radius in these galaxies (Figure \ref{fig:reff_evo}).}
\label{fig:slope_ratios}
\end{figure}

\begin{figure}
\centering
\includegraphics[width=\columnwidth]{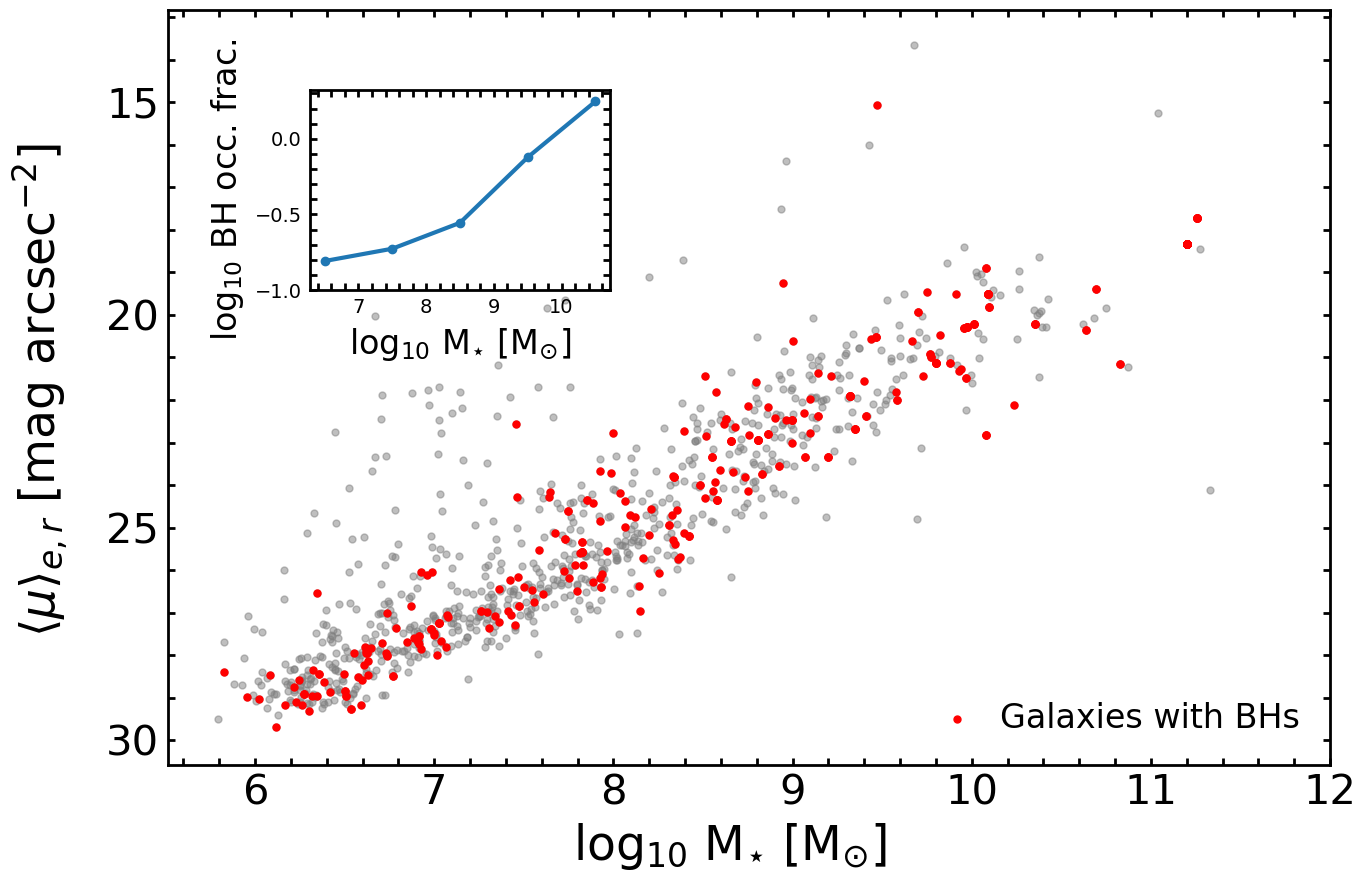}
\caption{Main panel: Surface brightness vs stellar mass for all galaxies (grey) and galaxies hosting a supermassive black hole (red). Inset: The occupation fraction of BHs in galaxies as a function of stellar mass. A small minority of dwarf galaxies in \texttt{NewHorizon} have central BHs, and that those that do, do not occupy a preferential position in the locus.}
\label{fig:agn}
\end{figure}

We now consider whether a similar process operates in the dwarf galaxy regime. While galaxies in all mass bins show similar behaviour, we first demonstrate the impact of SN feedback graphically, by showing the evolution of the stacked stellar density profiles in the high mass bin. Figure \ref{fig:prof} shows how the stacked stellar density profiles of all stars and that of young stars (ages < 100 Myr), which trace the star-forming gas, evolve across the epoch of divergence in SN feedback energy.

The more rapid star formation in lower locus galaxies at early epochs causes both the normalisation and the slope of the stacked young-star density profiles to decrease faster than in their upper locus counterparts, as star-forming gas is displaced from the centre towards the outskirts at a faster rate. As the systems evolve, progressive generations of young stars with flatter profiles will then drive a flattening of the overall stellar profiles. To quantify this, we fit a straight line to the inner 0.5 R$_{\rm{eff}}$ of each density profile (where the profile is roughly linear) and calculate the ratio in the slopes of the density profiles between the young stars and the total stellar population. As shown in Figure \ref{fig:slope_ratios}, the ratio of the slopes between young and all stars is smaller in lower locus galaxies for most of cosmic time (regardless of the mass bin being considered). This indicates that the density profiles of the young stars flatten faster in lower locus galaxies than in their upper locus counterparts. This, in turn, flattens their total stellar profiles faster, which is reflected in lower locus galaxies growing their effective radii more rapidly than their upper locus counterparts (Figure \ref{fig:reff_evo}). 

While initially the enhanced star formation in lower locus galaxies counteracts their larger effective radii and keeps their surface brightnesses similar to their upper locus counterparts, as star formation activity decreases in the lower locus (Figure \ref{fig:snevo}), the two populations diverge rapidly in surface brightness (Figure \ref{fig:surface brightnessevo}). While this trend is the same in all mass bins, the divergence in the level of star formation occurs later at higher masses (Figure \ref{fig:snevo}), and so the divergence in surface brightness also takes place at later epochs. Lower locus galaxies can be thought of as somewhat rarer events \citep[in the peak background split sense, e.g.][]{Schmidt2013} that start to form stars `too early' for the depth of their potential wells to counteract the energy release of the first round of supernovae in these galaxies. 

While the role of SN feedback in initiating the surface brightness divergence is clear, we conclude this section by considering whether another internal process, AGN feedback, may contribute to this process. In Figure \ref{fig:agn} (main panel) we show the location of galaxies that have central super-massive BHs (red points), compared to galaxies that do not (grey points). In the inset, we show the occupation fraction of central BHs in galaxies (defined as a galaxy having at least one BH within 1 R$_{\rm{eff}}$) as a function of stellar mass. In the low-mass regime, which is our mass range of interest, only a small fraction of galaxies actually host BHs (the BH occupation fractions are around 10 per cent for M${_\star}$ $\sim$ 10$^8$ M$_{\odot}$). It is worth noting that BHs that do exist in dwarf galaxies exhibit very little growth \cite[e.g.][]{Volonteri2020,Dubois2020}. Finally, galaxies with BHs do not occupy a preferential position in the locus. Taken together, this indicates that only a small minority of dwarf galaxies in \texttt{NewHorizon} have central BHs, and that those that do, do not occupy a preferential location in the locus and show very little growth (and therefore little potential for AGN feedback). Thus AGN feedback does not play a role in determining the position of galaxies in the locus.

%.........................................................

\subsubsection{A cosmological trigger for the stellar assembly bias: higher local dark matter density driving higher gas accretion rates}

In the previous section, we showed that galaxies that lower locus galaxies (i.e. those that end up with lower surface brightnesses at late epochs) are those that experience an earlier phase of more intense star formation. While this earlier stellar assembly is ultimately responsible for driving the initial divergence in surface brightness, we now explore the factors that are responsible for causing this bias in the first place. A faster stellar assembly implies a greater availability of gas. Thus, we explore the local environmental densities around lower and upper locus galaxies, at the epochs where the SN feedback energy diverges, since denser local environments will likely induce larger inflows of gas which can trigger stronger star formation.

In Table \ref{tab:top_bottom_ratio}, we show the local DM density, on scales of 15 R$_{vir}$ around each galaxy's halo, just before the epoch at which the SN feedback divergence takes places (lookback times of $\sim$11.9 Gyrs, $\sim$11.5 Gyrs and $\sim$11 Gyrs for the low, intermediate and high mass bins respectively). In all mass bins, lower locus galaxies reside in regions of higher average dark-matter density. The local DM densities in the lower locus galaxies are elevated by 22, 58 and 57 per cent in the low, intermediate and high mass bins respectively. %We also look at the rate that of gas inflow into the 15 R$_{vir}$ region, we calculate this as the gas mass increase divided by the lookback time between timesteps. To ensure that the result is not anomalous we take the median value of multiple timesteps over a window of 45 Myr. 
The gas inflow rates into this 15 R$_{vir}$ region are, in turn, elevated by 55, 48 and 63 per cent respectively. Note that these trends remain unchanged if we calculate local DM density on different spatial scales around the galaxies (e.g. 10 R$_{vir}$, 20 R$_{vir}$ etc.).

Lower locus galaxies are, therefore, born in regions of higher DM density, which enables them to access gas at a faster rate at earlier epochs. This faster gas accretion rate then drives the faster stellar assembly in the lower locus systems, leading to increased SN feedback and a faster increase in their effective radii. It is worth noting that, the idea that the density of the local environment in which dwarf galaxies form may imprint itself on their star formation histories, has been previously postulated for galaxies in the local group \citep[e.g.][]{Gallart2015}. Our work shows that this is a general property of dwarf galaxies and not a specific feature of dwarfs in the local group.

% \begin{center}
% \begin{table}
% \centering
% \begin{tabular}{| c | c | c | c |}
% \hline
% \hline
% Lower:upper locus ratio & Low mass & Int. mass & High mass\\
% \hline
% \hline
% DM density  & 1.22 & 1.58 & 1.57\\
% Gas inflow rate & 1.55 & 1.48 & 1.63\\
% \end{tabular}
% \caption{Ratio of the median dark matter density within 15 R$_{vir}$ (first row) and the median gas inflow rate (second row) between lower and upper locus galaxies in different mass bins.}
% \label{tab:top_bottom_ratio}
% \end{table}
% \end{center}

\begin{center}
\begin{table}
\centering
\begin{tabular}{| c | c | c | c |}
\hline
\hline
Lower:upper locus ratio & Low mass & Int. mass & High mass\\
\hline
\hline
DM density  & 1.22$\pm$0.16 & 1.58$\pm$0.22 & 1.57$\pm$0.28\\
Gas inflow rate & 1.55$\pm$0.05 & 1.48$\pm$0.06 & 1.63$\pm$0.08\\
\end{tabular}
\caption{Ratio of the median dark matter density within 15 R$_{vir}$ (first row) and the median gas inflow rate (second row) between lower and upper locus galaxies in different mass bins with associated standard errors.}
\label{tab:top_bottom_ratio}
\end{table}
\end{center}

Finally, it is worth exploring whether the higher DM densities experienced by lower locus galaxies are due to them residing in special locations in the cosmic web, or whether lower locus galaxies are simply sampling higher values of the local dark matter density distribution. We use the skeleton, described in Section \ref{sec:local_environment}, to check whether the locations of lower locus galaxies vary systematically from their upper locus counterparts. However, we find that both populations show similar distances to both the nearest nodes and filaments, indicating that, on average, the positions of upper and lower locus galaxies in the cosmic web are very similar. Note that, since our sample only includes galaxies that survived accretion on to more massive galaxies down to $z=0.25$, it may be potentially biased towards objects that are born relatively further from the densest regions of the skeleton, since such systems will have a lower probability of merging with a larger galaxy \citep[e.g.][]{Borzyszkowski2017,Musso2018}. %The trends seen here may not, therefore, be relevant to the general dwarf population at high redshift.

%{\color{red} It should be noted however that the statistics for this are poor as the galaxy population sizes are small, particularly at high redshift and low mass. Also, due to these galaxies existing at $z=0.25$ there is a selection bias towards galaxies that must have formed away from the skeleton, else they would have merged with a larger galaxy, this may dampen trends that would normally be seen. Our results therefore, are not a statement that dwarf galaxy formation does not correlate with the cosmic web but that dwarf galaxies that survive are more influenced by local environment.}

Our analysis has shown that the differences between the upper and locus galaxies are driven by the fact that their progenitors sample different parts of the local dark matter density distribution in the Universe at early epochs. \textit{The stellar assembly bias which drives the divergence in galaxy surface brightness at fixed stellar mass, and therefore the width of the main locus at $z=0.25$ (Figure \ref{fig:mvssurface brightness}), essentially has a purely cosmological trigger.} 

%.........................................................

\subsection{Tidal perturbations and ram pressure: external processes that influence dwarf galaxy evolution at late epochs}

At late epochs, the level of star formation in lower locus galaxies subsides and SN feedback activity in these systems falls below the levels seen in their upper locus counterparts (Figure \ref{fig:snevo}). Nevertheless, the effective radii of lower locus systems keeps growing faster than in their upper locus counterparts (Figure \ref{fig:reff_evo}). The continued divergence in surface brightness between the two populations at late epochs (Figure \ref{fig:surface brightnessevo}) is thus driven both by lower locus galaxies forming stars at a slower rate and growing their effective radii more rapidly than their upper locus counterparts. Most importantly, since the star formation activity in the lower locus galaxies is now less intense, the evolution in their radii must be driven by external processes and not by internal mechanisms like SN feedback.

The two key external processes that can influence galaxy evolution are tidal perturbations (including mergers) and ram pressure. Strong tidal perturbations from events like close fly-bys and mergers can trigger star formation. Sustained tidal perturbations over a long period of time can heat stars and gas, puffing up galaxies and reducing their star formation rate \citep{Martin2019}. Sustained ram pressure typically acts to strip the internal gas reservoir of the galaxy as it transits through a dense medium, reducing its star formation rate \citep[e.g.][]{Hester2006}. 

\begin{figure}
\centering
\includegraphics[width=\columnwidth]{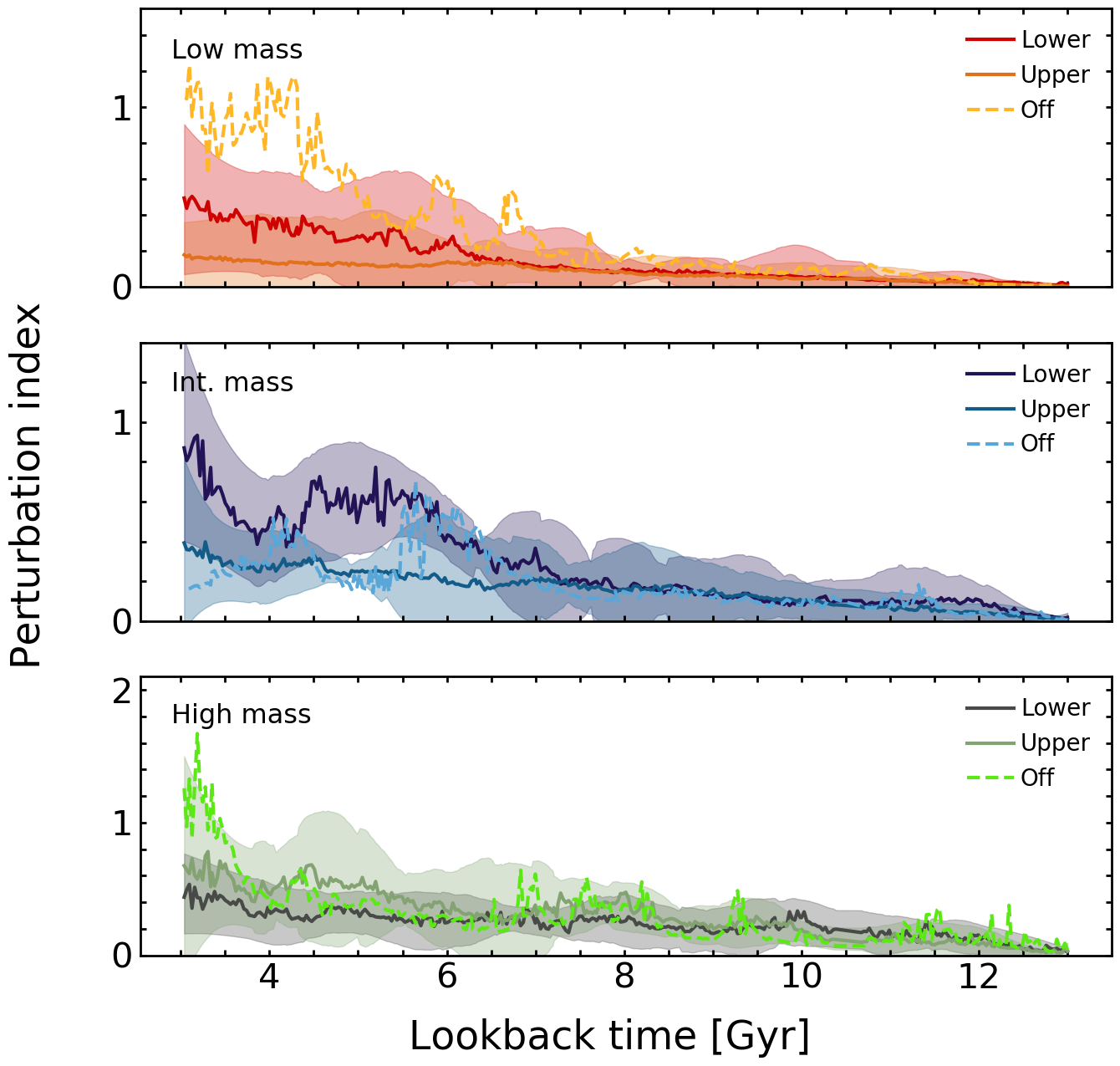}
\caption{Evolution of the median perturbation index (PI) with look-back time. The shaded regions show the associated uncertainties. The top, middle and bottom panels represent the low (10$^{6.5}$ M$_{\odot}$ $\geq$ M$_{\star}$ $>$ 10$^{7.5}$ M$_\odot$), intermediate (10$^{7.5}$ M$_\odot$ $\geq$ M$_{\star}$ $>$ 10$^{8.5}$ M$_{\odot}$) and high (10$^{8.5}$ M$_{\odot}$ $\geq$ M$_{\star}$ $>$ 10$^{9.5}$ M$_{\odot}$) mass bins respectively. The colour coding indicates the upper, lower and off locus galaxy populations. Recall that lower locus galaxies represent the population that is fainter in surface brightness at $z=0.25$, while the upper locus galaxies represent their brighter counterparts. Upper and lower locus galaxies exhibit similar PI evolution at early epochs, with the PI experienced by lower locus galaxies increasing at late epochs. Off-locus objects show large, recent increases in PI, which indicates that their change in surface brightness is influenced by a recent interaction.}
\label{fig:PI}
\end{figure}

\begin{figure*}
\centering
\includegraphics[width=\textwidth]{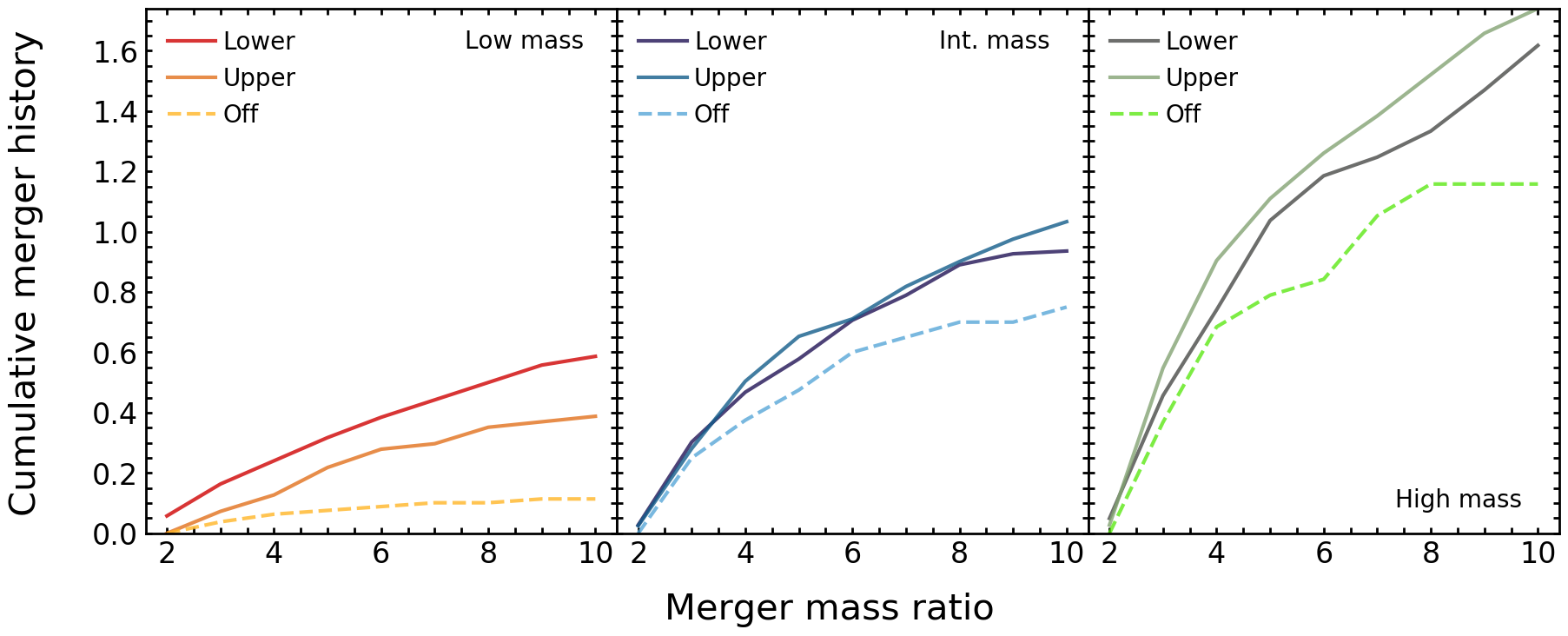}
\caption{Cumulative merger history for galaxies in the different zones indicated in Figure \ref{fig:mvssurface brightness}. The left, centre and right panels represent the low, intermediate and high mass bins respectively. The colour coding indicates the upper, lower and off locus galaxy populations. This figure presents the average number of mergers experienced by a galaxy over its lifetime, with mass ratios less than or equal to the value shown on the x-axis.} 
\label{fig:merger}
\end{figure*}

\begin{figure}
\centering
\includegraphics[width=\columnwidth]{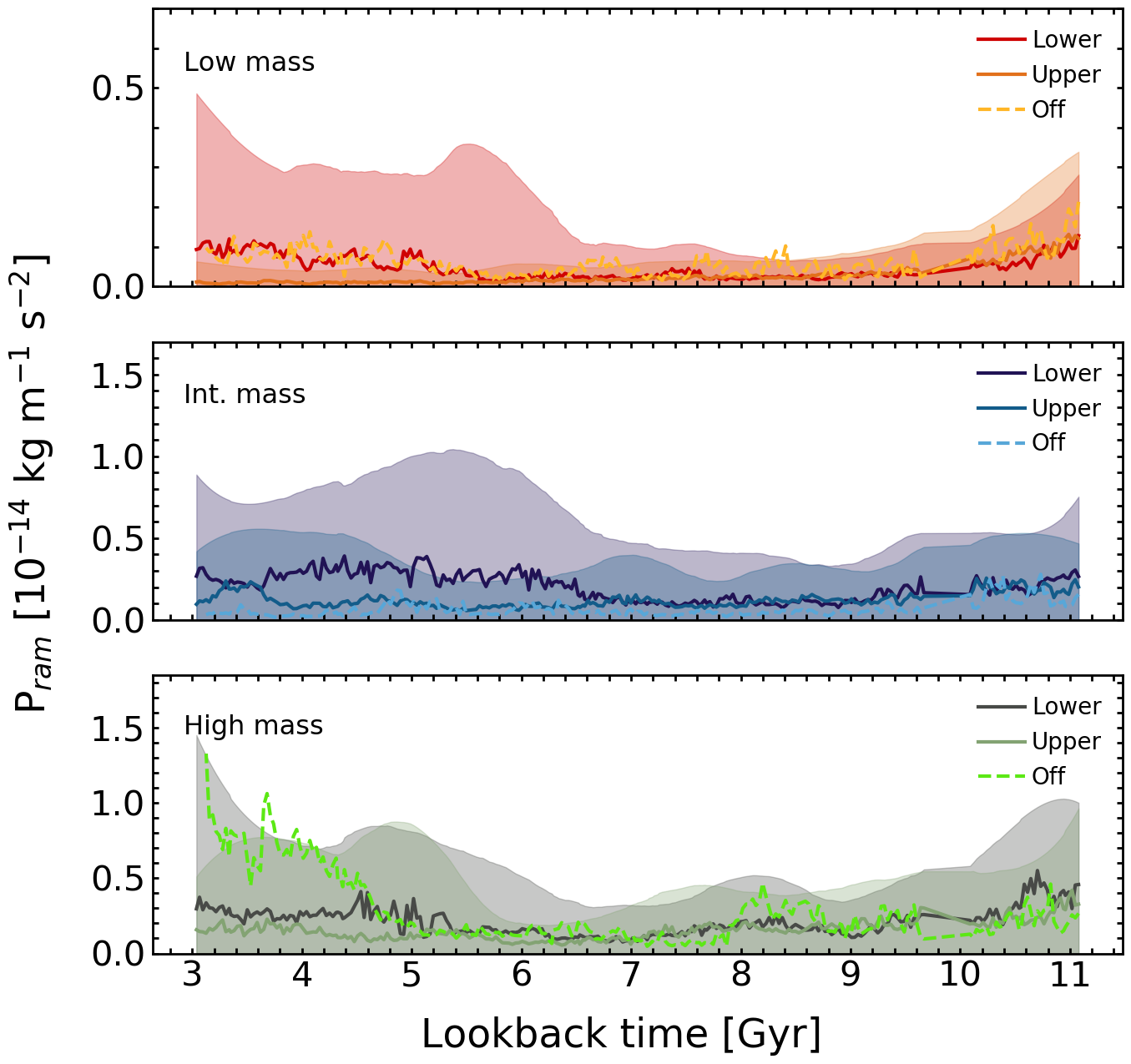}
\caption{Evolution of the median ram pressure with look-back time. The shaded regions show the associated uncertainties. The top, middle and bottom panels represent the low (10$^{6.5}$ M$_{\odot}$ $\geq$ M$_{\star}$ $>$ 10$^{7.5}$ M$_\odot$), intermediate (10$^{7.5}$ M$_\odot$ $\geq$ M$_{\star}$ $>$ 10$^{8.5}$ M$_{\odot}$) and high (10$^{8.5}$ M$_{\odot}$ $\geq$ M$_{\star}$ $>$ 10$^{9.5}$ M$_{\odot}$) mass bins respectively. The colour coding indicates the upper, lower and off locus galaxy populations. Recall that lower locus galaxies represent the population that is fainter in surface brightness at $z=0.25$, while the upper locus galaxies represent their brighter counterparts. In all mass bins, galaxies in different zones of the locus experience similar values of ram pressure over cosmic time.}
\label{fig:ram}
\end{figure}

Following \citet{Choi2018} and \citet{Martin2019} we define a dimensionless `perturbation index' (PI) that quantifies the cumulative, ambient tidal field around a galaxy:

\begin{equation}
\label{eqn:PI}
PI = \sum_{i}\left ( \frac{M_{i}}{M_{gal}}\right ) \left ( \frac{R_{\mathrm{eff}}}{D_{i}} \right )^{3},
\end{equation}

\noindent where $M_{\rm{gal}}$ is the stellar mass of the galaxy in question and $R_{\mathrm{eff}}$ is its effective radius, $M_{i}$ is the stellar mass of the $i$th perturbing galaxy and $D_{i}$ is the distance from the $i$th perturbing galaxy. We consider all perturbing galaxies within 3 Mpc of the object in question.

Figure \ref{fig:PI} shows the evolution of the median PI with look-back time. We find that in all mass bins, galaxies that reside in different zones of the locus experience very similar tidal perturbations at high redshift. Therefore, perturbations caused by the ambient tidal field do not contribute to the early divergence in surface brightness. However, in the low and intermediate mass bins, lower locus galaxies show higher values of PI at late epochs (likely driven by the fact that these galaxies are born, and remain in, regions of higher density) which contribute to the divergence in their effective radii and surface brightnesses at late epochs. Interestingly, the trend is reversed in the high mass bin, indicating that the surface brightness divergence in this regime is driven primarily by the higher levels of star formation in the upper locus population.

Note that, while the PI includes mergers (i.e. events where the two galaxies eventually coalesce), it is worth looking separately at the merger history of galaxies, since plotting the median PI will dampen the effect of individual events, like mergers, which could cause a strong, transient change in this parameter. While mergers will increase the stellar mass of the system and can drive an increase in its radius, the smooth change in both stellar mass (Figure \ref{fig:mass_evo}) and effective radius (Figure \ref{fig:reff_evo}) suggests that mergers, which are stochastic and rare events \citep[e.g.][]{Darg2010,Uzeirbegovic2020}, are unlikely to be driving this evolution. We confirm this in Figure \ref{fig:merger} by exploring the cumulative merger histories of galaxies, for significant merging events which have mass ratios less than 10:1 (which is the mass ratio range where mergers typically produce measurable size and morphological change, e.g. \citet{Martin2018b}). This figure shows the average number of mergers experienced by a galaxy since the beginning of the simulation, with mass ratios less than or equal to the value shown on the x-axis. For example, a lower locus galaxy in the low mass bin (left panel) undergoes, on average, $\sim$0.6 mergers with mass ratios greater than 1:10 at $z>0.25$, while a lower locus galaxy in the intermediate mass bin undergoes, on average, $\sim$0.95 mergers in the same time period. 

We find that the cumulative merger histories of galaxies in the lower locus do not show a strong preference towards higher merger fractions over cosmic time in any mass bin. In addition, the number of mergers experienced by galaxies (in all zones of the locus) are indeed too small to explain the smooth change observed in stellar mass and effective radius. This indicates that individual mergers do not influence the eventual position of a galaxy on the locus.

Next, we consider ram pressure, which is capable of removing gas from galaxies and reducing their star formation rates, particularly in high-density environments \citep[e.g.][]{Gunn1972}. While \texttt{NewHorizon} does not include the richest environments like galaxy clusters, ram pressure can still play a significant role in the evolution of low-mass galaxies, which are typically much more susceptible to gas stripping due to their shallow gravitational potential wells \citep[e.g.][]{Vollmer2001,Martin2019}. We measure the ram pressure exerted on a galaxy by the local medium as follows:

\begin{equation}
\label{eqn:ram-pressure}
P_{ram}=\rho_{\mathrm{IGM}} v_{\mathrm{gal}}^{2},
\end{equation}

\noindent where $v_{gal}$ is the velocity of the galaxy relative to the bulk velocity of the surrounding medium and $\rho_{\mathrm{IGM}}$ is the mean gas density of the surrounding medium within 10 times the maximum extent of the stellar distribution of the galaxy.

Figure \ref{fig:ram} shows the evolution of the median ram pressure with look-back time. In a similar vein to the PI, the ram pressure experienced by upper and lower locus galaxies is similar at large look-back times and therefore does not contribute to the surface brightness divergence at early epochs. However, the median ram pressure experienced by lower locus galaxies at late epochs is elevated compared to that in their upper-locus counterparts (although the difference is less pronounced than what is found for the PI). This is again likely to be driven by the fact that lower-locus galaxies are born, and remain in, regions of higher density. The higher ram pressure at late epochs likely assists in the faster depletion of gas in lower locus galaxies, reducing their star formation activity (which is reflected in the decrease in the SN feedback energy seen in Figure \ref{fig:snevo}). This, in turn, helps drive the divergence in surface brightness between the two populations at late epochs. 

%.........................................................

\section{Off-locus galaxies: transient, tidally-induced starbursts}
\label{sec:off_locus_section}

We complete our study by exploring the formation mechanisms of galaxies that depart strongly from the locus towards higher surface brightnesses. As shown in Figure \ref{fig:mvssurface brightness}, a minority of galaxies reside `off-locus', with the fraction of off-locus objects being 22, 12 and 10 per cent in the low, intermediate and high mass bins respectively (see Table \ref{tab:off_locus}). In particular, as noted in Section \ref{sec:galaxy_prop}, galaxies that scatter off the main locus make up the majority of low-mass galaxies that are visible in past wide-area surveys like the SDSS. It is, therefore, important to understand how these systems form and their connection to the majority of the galaxy population which resides on the locus itself.  

Figure \ref{fig:surface brightnessevo} suggests that the off-locus population diverges in surface brightness at late look-back times, with the divergence being most pronounced in the low-mass population. The epoch of divergence coincides with an increase in both the PI (Figure \ref{fig:PI}) and the SN feedback energy i.e. a starburst (Figure \ref{fig:snevo}). We first check whether the gas inflow rates in the off-locus galaxies show an increase around the time they begin to diverge from the population that resides in the locus, possibly due to abruptly entering a high-density environment. However, we find that the average gas inflow rates prior to the starburst are not anomalous and an analysis of the skeleton indicates that the off-locus galaxies do not abruptly enter high-density regions of the cosmic web. 

This indicates that the off-locus galaxies depart from the main locus due to recent, tidally-induced starbursts which are essentially stochastic in nature. The epoch at which the population diverges in surface brightness (see Figure \ref{fig:surface brightnessevo}) indicates that the star formation episodes in these systems are around 1-4 Gyrs old (dependent on the mass bin). Note that this recent starburst episode is not special. While galaxies are likely to have had multiple starburst episodes in their lifetimes, the new stars produced in those episodes have since aged and faded away. The off-locus population is essentially tracing the most recent starburst episode whose signature remains visible in the $r$-band filter.  

\begin{figure}
\center
\includegraphics[width=\columnwidth]{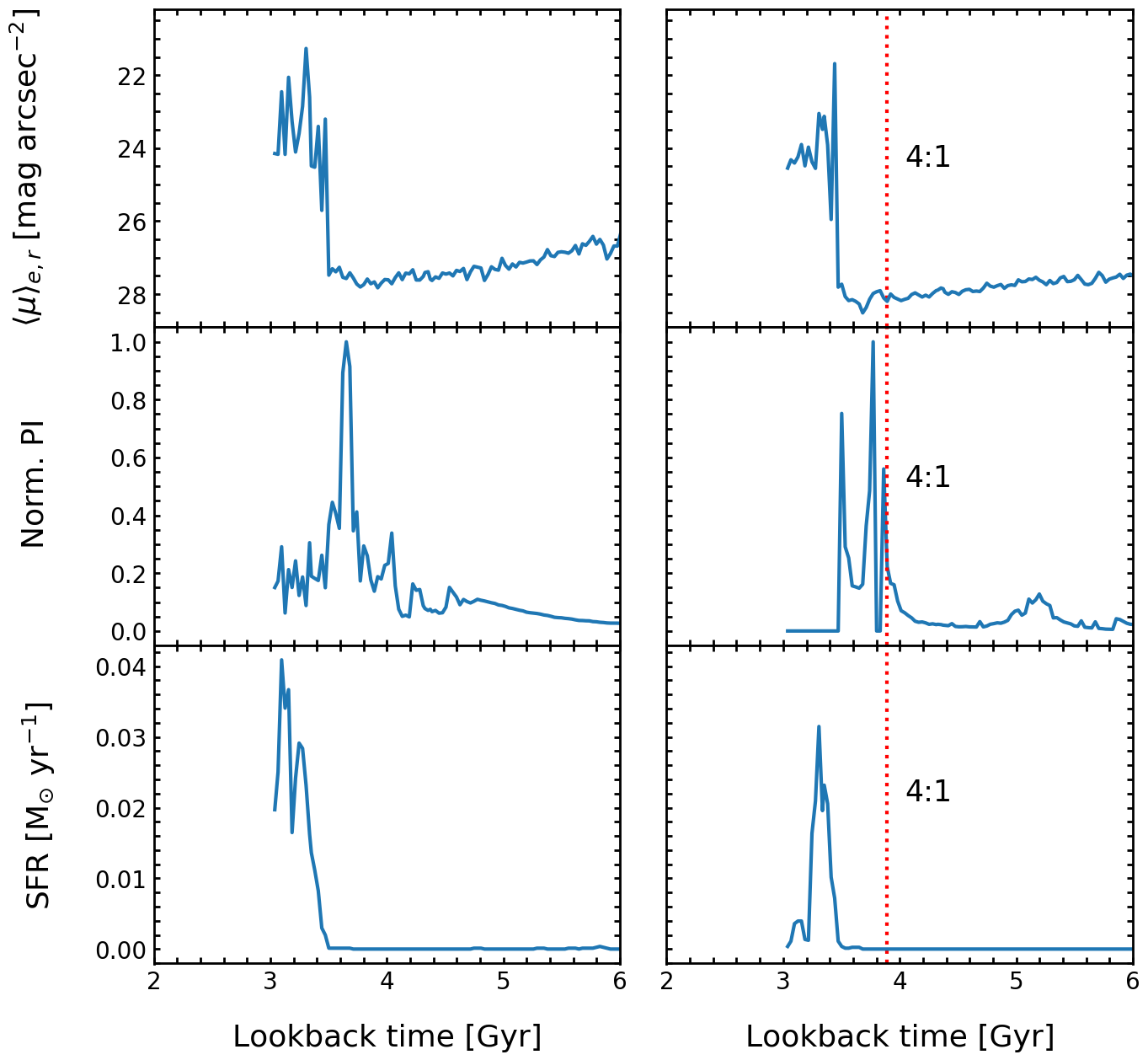}
\caption{Top row: Evolution of surface brightness with look-back time for two off-locus galaxies in the high mass bin. Bottom row: The corresponding evolution of the perturbation index with look-back time. Vertical dotted red lines indicate the times at which merger events take place. While the galaxy in the right-hand column exhibits a rapid increase in surface brightness that is coincident with a merger with mass ratio 4:1, the increase in surface brightness of the galaxy in the left-hand column does not correspond to a merger event and is therefore driven by a fly-by.}
\label{fig:PIofflocus}
\end{figure}

\begin{center}
\begin{table}
\centering
\begin{tabular}{| c | c | c | c |}
\hline
\hline
 & Low mass & Int. mass & High mass\\
\hline
\hline
Off-locus fraction  & 22$\%$ & 12$\%$ & 10$\%$\\\hline
Fly-by induced & 92$\%$ & 48$\%$ & 38$\%$\\
Major-merger induced & 4$\%$ & 26$\%$ & 12$\%$\\
Minor-merger induced & 4$\%$ & 26$\%$ & 50$\%$\\
\end{tabular}
\caption{The fraction of off-locus galaxies (first row) and the mechanisms that produce off-locus systems (other rows) in different mass bins. For example, in the intermediate mass bin, 12 per cent of galaxies are found off-locus. Of these 48 per cent are fly-by induced while 26 per cent are induced by major mergers (mass ratios $<$ 4:1) and the rest are triggered by minor mergers (mass ratios $>$ 4:1).}
\label{tab:off_locus}
\end{table}
\end{center}

Since the increase in PI could either be driven by fly-bys i.e. purely tidal events where there is a rapid change in PI without a merger or by major (mass ratios $<$ 4:1) or minor (mass ratios $>$ 4:1) mergers, it is instructive to consider which mechanism triggers the starburst that produces off-locus galaxies in different zones of the locus. For each off-locus galaxy, we consider their merger histories in conjunction with the evolution of their PI values and compare this to their surface brightness evolution. Figure \ref{fig:PIofflocus} shows examples of two cases. The right-hand column shows a system that is driven off-locus (due to the rapid brightening of the surface brightness, see top panel) by an increase in PI that coincides with a recent merger with a mass ratio of 4:1. The look-back time of the merger is indicated by the red dotted line. The left-hand column shows a case where the surface brightness is driven by a fly-by because the change in the perturbation index does not correspond with a recent merger. 

We visually inspect these plots for every off-locus galaxy, in order to ascertain the principal driver of the starburst that moves them off the locus. Table \ref{tab:off_locus} summarises the drivers of the starbursts in the off-locus objects. The impact of fly-bys is greatest in the low-mass regime, where 92 per cent of the off-locus systems are fly-by induced. Fly-bys remain important in all mass regimes, producing 48 and 38 per cent of off-locus systems in the intermediate and high mass bins respectively. However, mergers become more important at higher stellar masses, accounting for 62 per cent of off-locus objects in the high mass bin. The progressively higher minor-merger-induced off-locus fraction at higher stellar masses is due to the shape of the galaxy mass function. Since there are fewer galaxies at high stellar masses \citep[e.g.][]{Bell2003}, equal mass mergers between massive galaxies become rarer in this mass regime. 

Our analysis indicates that the recent starbursts that produce off-locus systems are largely induced by tidal perturbations that trigger the existing gas reservoirs in these galaxies, rather than being due to enhanced gas accretion from the cosmic web. As galaxies increase in stellar mass and their gravitational potential wells become deeper, a larger perturbation, via a merger rather than simply a fly-by, is required to trigger this starburst.

It is important to note that off-locus galaxies are clearly atypical of the general galaxy population because they have anomalously high levels of star formation. As has been noted before in Section \ref{sec:galaxy_prop}, in the M$_{\star}$ $<$ 10$^{9}$ M$_{\odot}$ regime, off-locus galaxies overwhelmingly dominate the observed galaxy population in past wide-area surveys like the (standard-depth) SDSS (see dotted red lines in Figure \ref{fig:locus_comparison}). Therefore, conclusions about the general galaxy population cannot be drawn using the subset of off-locus objects because they are likely to be highly unrepresentative of the overall galaxy populations in these mass regimes. 

%.........................................................

\section{Summary}
\label{sec:summary}

Our statistical comprehension of galaxy evolution is driven by objects that are brighter than the surface brightness limits of wide-area surveys. Given the depth of past wide surveys, our understanding of how the observable Universe evolves has been predicated largely on bright galaxies. For example, while past benchmark surveys like the SDSS become highly incomplete at effective surface brightnesses of around 24 mag arcsec$^{-2}$, low-surface brightness galaxies (LSBGs) that are fainter than these thresholds overwhelmingly dominate the galaxy number density \citep[e.g.][]{Martin2019}. Understanding the evolution of the LSBG population is therefore key to a complete comprehension of galaxy evolution. Here, we have used the high resolution \texttt{NewHorizon} cosmological simulation to probe the origin and evolution of the LSBG population in the low-mass regime, where most recent observational studies are focussed. In particular, we have explored why, at a given stellar mass, LSBGs span the large observed range in surface brightness. 

The \texttt{NewHorizon} galaxy population occupies a well-defined locus in the surface brightness vs. stellar mass plane, with a large spread of $\sim3$ mag arcsec$^{-2}$. A minority of galaxies depart strongly from this locus towards higher surface brightnesses. The fraction of these off-locus systems is $\sim$20 per cent in the low mass bin (10$^{6.5}$ M$_{\odot}$ $<$ M$_{\star}$ $<$ 10$^{7.5}$ M$_{\odot}$) and drops to $\sim$10 per cent at higher stellar masses (10$^{8.5}$ M$_{\odot}$ $<$  M$_{\star}$ $<$ 10$^{9.5}$ M$_{\odot}$). The predicted position of \texttt{NewHorizon} galaxies in the surface brightness vs. stellar mass plane is in good agreement with observations from the relatively deep SDSS Stripe 82 survey. This agreement suggests that, while prescriptions used in \texttt{NewHorizon} to describe baryonic physics (e.g. SN feedback) are largely tuned to higher mass galaxies, they also offer a good representation of these processes in the  low-mass regime. 

A large number of LSBGs in \texttt{NewHorizon}, particularly galaxies that exist in the upper locus in the low and intermediate mass bins, have surface brightnesses and effective radii that make them consistent with the definition of `ultra-diffuse galaxies' (UDGs) in the current observational literature. These dwarfs form naturally in the standard model, through a combination of SN feedback, tidal perturbations and ram pressure as described in Section \ref{sec:galaxy_evol} and are, therefore, a normal component of the dwarf galaxy population (albeit one that is not readily detectable using past and current datasets). Since \texttt{NewHorizon} does not contain very high-density environments (i.e. clusters), a clear prdiction is that UDGs form ubiquitously in low-density environments like groups and the field and will be routinely visible in new and future deep-wide surveys like the HSC-SSP and LSST. 

To understand the processes that determine the eventual surface brightness of dwarf galaxies at late epochs, we have split the locus into three zones - the lower (fainter) and upper (brighter) halves and the population of off-locus galaxies that scatter towards very high surface brightnesses beyond the upper locus. We have then studied, in detail, the formation histories of galaxies in these different zones, in order to identify the processes that create the large observed spread in surface brightnesses in the dwarf galaxy population. 

Regardless of the mass regime being considered, galaxies in the lower locus, i.e. those that end up with fainter surface brightnesses at late epochs, are born in denser regions of the Universe, which results in faster gas accretion and more intense star formation at high redshift. %Lower locus galaxies do not appear to exhibit a special location within the cosmic web, compared to upper locus galaxies, and are simply objects that sample higher values of the DM density distribution at early epochs. 
The more intense star formation in these systems leads to stronger supernova feedback. This flattens gas profiles at a faster rate which, in turn, creates shallower stellar profiles more rapidly. As star formation subsides in the lower locus, their late epoch evolution is dominated by external processes like tidal perturbations and ram pressure. Since lower locus galaxies are born in higher density environments, they remain in relatively denser environments over their lifetimes. The higher tidal perturbations they experience as a result continue to increase their effective radii at a faster rate than their upper locus counterparts. In a similar vein, the higher ram pressure they experience acts to accelerate gas depletion and reduce their star formation rates more quickly than on the upper locus. The lower locus systems thus diverge strongly from their upper locus counterparts at late epochs, both due to the fact that they are now more diffuse and because they have less star formation.

Finally, we have studied the processes that drive the formation of off-locus galaxies that deviate strongly from the main locus as they experience a recent, rapid increase in surface brightness. This increase is not driven by an uptick in the gas inflow rate but coincides with an increase in the perturbation index and a coincident rise in the star formation activity. This indicates that an interaction triggers a starburst which moves the galaxy away from the main locus. In low-mass systems (M$_{\star}$ $<$ 10$^{7.5}$ M$_{\odot}$) this starburst is triggered primarily by purely tidal events (i.e. fly-bys), while at high stellar masses (M$_{\star}$ $>$ 10$^{8.5}$ M$_{\odot}$) it is primarily induced by mergers. It is worth noting that this off-locus population makes up the bulk of the observable low-mass systems (M$_{\star}$ $<$ 10$^{8.5}$ M$_{\odot}$) in past benchmark wide-area surveys like the SDSS. However, since these systems are clearly anomalous in terms of their star formation activity, conclusions about the general galaxy population cannot be drawn using these galaxies. 

Together with the study presented in \citet{Martin2019}, the analysis in this paper offers a theoretical counterpart to new and future deep-wide surveys like the HSC-SSP and LSST. As noted earlier, these surveys will offer galaxy completeness down to at least  M$_{\star}$ $\sim$ 10$^7$ M$_{\star}$ in the nearby Universe, providing a revolutionary increase in discovery space. In future papers, we will compare our theoretical studies to data from such surveys to explore LSBG formation and constrain the physics of galaxy evolution, in detail, in the low-surface-brightness regime.  

%.........................................................

\section*{Acknowledgements}
We are grateful to the anonymous referee for many constructive comments that improved the quality of the original manuscript. We warmly thank Rob Crain and Johan Knapen for many interesting discussions. RAJ, GM and SK acknowledge support from the STFC [ST/R504786/1, ST/N504105/1, ST/S00615X/1]. SK acknowledges a Senior Research Fellowship from Worcester College Oxford. The research of AS and JD is supported by the Beecroft Trust and STFC.  Some of the numerical work made use of the DiRAC Data Intensive service at Leicester, operated by the University of Leicester IT Services, which forms part of the STFC DiRAC HPC Facility (www.dirac.ac.uk). The equipment was funded by BEIS capital funding via STFC capital grants ST/K000373/1 and ST/R002363/1 and STFC DiRAC Operations grant ST/R001014/1. DiRAC is part of the National e-Infrastructure. TK was supported in part by the National Research Foundation of Korea (NRF-2017R1A5A1070354 and NRF-2020R1C1C100707911) and in part by the Yonsei University Future-leading Research Initiative (RMS2-2019-22-0216). This research has used the DiRAC facility, jointly funded by the STFC and the Large Facilities Capital Fund of BIS, and has been partially supported by grant Segal ANR- 19-CE31-0017 of the French ANR. This work was granted access to the HPC resources of CINES under the allocations 2013047012, 2014047012 and 2015047012 made by GENCI. This work was granted access to the high-performance computing resources of CINES under the allocations c2016047637 and A0020407637 from GENCI, and KISTI (KSC-2017-G2-0003). Large data transfer was supported by KREONET, which is managed and operated by KISTI. This work has made use of the Horizon cluster on which the simulation was post-processed, hosted by the Institut d'Astrophysique de Paris. We thank Stephane Rouberol for running it smoothly for us. S.K.Y. acknowledges support from the Korean National Research Foundation (NRF-2020R1A2C3003769). This work is part of the Horizon-UK project. 

%.........................................................

\section*{Data Availabilty}

The data that underpins the analysis in this paper were provided by the Horizon-AGN and \texttt{NewHorizon} collaborations. The data will be shared on request to the corresponding author, with the permission of the Horizon-AGN and \texttt{NewHorizon} collaborations.  

%.........................................................

%%%%%%%%%%%%%%%%%%%%%%%%%%%%%%%%%%%%%%%%%%%%%%%%%%



\bibliographystyle{mnras}
\bibliography{bib} % if your bibtex file is called example.bib

\bsp	% typesetting comment
\label{lastpage}
\end{document}